\documentclass[12pt,a4paper]{article}

\usepackage[latin1]{inputenc}
\usepackage{amsmath}
\usepackage{nccmath}
\usepackage{amsfonts}
\usepackage{amssymb}
\usepackage{setspace}
\usepackage{tensor}

\usepackage{axodraw4j}
\usepackage{graphicx}
\usepackage{pstricks}
\usepackage{color}

\usepackage{array}

\usepackage{float}
\usepackage[margin=1cm]{caption}
\usepackage{minitoc}
\usepackage{makeidx}

\usepackage[backend=biber,style=authoryear,citestyle=authoryear-comp,sortcites=false,firstinits=true,maxcitenames=2,natbib=true]{biblatex}

\DeclareNameAlias{sortname}{last-first}
\usepackage{enumerate}
\usepackage{fancyhdr}
\usepackage[left=3.50cm, right=3.50cm, top=3.00cm, bottom=3.00cm]{geometry}
\usepackage{setspace}

\usepackage{caption}

\usepackage{lineno}

\author{S\'ebastien Rivat\thanks{Max Planck Institute for the History of Science, Berlin. Email: srivat@mpiwg-berlin.mpg.de}}
\title{Drawing Scales Apart: The Origins of Wilson's Conception of Effective Field Theories}

\usepackage{epigraph}
\begin{document}

\maketitle

\begin{abstract}
	This article traces the origins of Kenneth Wilson's conception of effective field theories (EFTs) in the 1960s. I argue that what really made the difference in Wilson's path to his first prototype of EFT are his long-standing pragmatic aspirations and methodological commitments. Wilson's primary interest was to work on mathematically interesting physical problems and he thought that progress could be made by treating them as if they could be analyzed in principle by a sufficiently powerful computer. The first point explains why he had no qualms about twisting the structure of field theories; the second why he divided the state-space of a toy model field theory into continuous slices by following a standard divide-and-conquer algorithmic strategy instead of working directly with a fully discretized and finite theory. I also show how Wilson's prototype bears the mark of these aspirations and commitments and clear up a few striking ironies along the way.
\end{abstract}


\normalsize

%\linenumbers
%\setlength\linenumbersep{1cm}

\section{Introduction}
\label{sec:Introduction}

Effective field theories (EFTs) have taken a central place in physics practice during the last decades. Many physicists today even believe that empirically successful theories are ultimately best treated as effective theories and thus, as a matter of design, as incomplete and non-fundamental theories. Yet little has been said so far about how this came to be. Historical studies on postwar physics have focused on the successful field-theoretic treatment of the electromagnetic interaction in the late 1940s, the most vocal anti-field-theoretic responses in the two following decades, the revival of field theory in the early 1970s, and the ultimate triumph of the Standard Model of particle physics by the end of the 1970s (e.g., \cite{pickering_constructing_1984,cushing_theory_1990,schweber_qed_1994,kaiser_drawing_2005,close_infinity_2013}). But for instance, no attention has been paid to the fact that the early form of the Standard Model was already treated as an EFT in 1973-74 (see, e.g., \cite{georgi_hierarchy_1974}).

The few historians who have looked at effective theories, such as Cao and Schweber (\cite*{cao_conceptual_1993}) and Schweber (\cite*{schweber_hacking_2015}), have also left crucial aspects out in their story. According to the standard account, the EFT program grew mainly out of Kenneth Wilson's work on renormalization group (RG) methods in the 1960-70s thanks to a fruitful cross-fertilization between particle and condensed matter physics.\footnote{See also Williams (\cite*[chap. 1]{williams_historical_2016}) for a preliminary historical account along these lines and Koberinski (\cite*{koberinski_mathematical_2021}) for similar remarks.} Wilson, however, formulated a first prototype of EFT before his encounter with condensed matter physics, and there has not yet been any detailed analysis of the origins of this early work. Likewise, the work of Steven Weinberg, Sydney Coleman, and many others on phenomenological Lagrangians in the late 1960s has not been much explored despite its decisive impact on Weinberg's (\cite*{weinberg_phenomenological_1979}) foundational article and, through this work, on many quarters of the EFT program (e.g., \cite[chap. 4]{donoghue_dynamics_1994}; \cite{van_kolck_effective_1999}).\footnote{Schweber acknowledges this omission and justifies it by pointing out that he is only interested in historical elements that transformed ``the meaning of quantum field theoretical descriptions" (\cite*[p. 68]{schweber_hacking_2015}). But he gives no ground to believe that the meaning of field theory was not also fundamentally altered by the rise of phenomenological Lagrangians. Note that Sakharov's treatment of General Relativity as an effective theory in (\cite*{sakharov_vacuum_1967}) and Coleman and E. Weinberg's (\cite*{coleman_radiative_1973}) work on effective potentials are closely related to this tradition and would also deserve special scrutiny.}
	
This paper is the first of a series aiming to tell a more comprehensive story of the early development of effective theories by focusing on Wilson's and Weinberg's works in the 1960s. During this period, Wilson and Weinberg both came to loosen in their own way the rules of ``conventional field theory" inherited from the successful perturbative formulation of quantum electrodynamics (QED), the quantum theory of the electromagnetic interaction. Wilson worked with discretized Hamiltonian models through his attempt to develop a new method to understand the high-energy structure of field theories. Weinberg worked with non-fundamental Lagrangian models with free parameters through his attempt to derive empirical quantities more systematically and efficiently. And despite their different trajectories, they each ended up with a prototype of EFT: namely, an approximate field theory including all the terms compatible with its principles and valid, by virtue of its mathematical structure, only within some limited regime.\footnote{This characterization, which I intend as a working definition here, is based on physicists' usual way of describing the most distinctive features of EFTs today (e.g., \cite[chap. 4]{donoghue_dynamics_1994}; \cite[chap. 1]{petrov_effective_2016}). For a detailed analysis of the general concept of effective theory across physics, see Rivat (\cite*[sec. 2-3]{rivat_effective_2021}). I also speak about ``principles" and not merely ``symmetries" to emphasize that the terms introduced are usually required to satisfy other assumptions, which are, strictly speaking, not symmetry principles (e.g., hermiticity, analyticity, locality).} 

The first fundamental point of contact between Wilson's and Weinberg's works appears, in fact, only in 1971 or so, when the most distinctive EFT feature of their prototypes, namely, that an EFT includes all the terms compatible with its principles, becomes fully clear to both of them.\footnote{There are, after all, many other types of approximate field theories valid only within some limited regime at the time, such as the static model for instance, as Chew and Low (\cite*{chew_effective-range_1956}) already made it clear.} I will thus end the story at this stage, in 1971, where nothing yet seems to prevent any kind of strong convergence afterward, and leave the question of whether Wilson's and Weinberg's works ultimately merged into a fully unified framework for future work. The important thing for now is that despite this point of contact, Wilson's and Weinberg's prototypes are still very different from each other in 1971, much more than one might expect if one believes that there is a fully unified EFT program that directly came out of Wilson's works, and this prompts two related questions: (i) What brought them to come up with such different prototypes? (ii) What really made the difference in each case? 

I will focus on Wilson's part of the story in this paper. Wilson formulated a first prototype of effective theory in his article ``Model Hamiltonians for Local Quantum Field Theory" in 1965. After reducing the state-space of a simple meson-nucleon model to a set of well-separated continuous slices, he worked out a sequence of low-energy effective models with increasingly large high-energy cut-offs by successively taking new high-energy slices into account. He also derived a sequence of couplings associated with these models---Wilson's first version of an RG equation. And although he only reached a fully-fledged conception of the RG and EFTs in 1971 or so, many of its elements were already apparent in this article: for instance, the key idea of separating degrees of freedom according to their relevance at different energy scales. I will thus follow Wilson's path to his first prototype, analyze his achievements in 1965 in relation to his later works, and briefly explore the way in which his interactions and concerns led him to his mature conception in the early 1970s. 

My central claim is that although Wilson's ``vehicle" was certainly field theory, as he puts it later on (\cite*[1/4, 17'58'']{wilson_interview_1991}), he was mainly driven toward effective theories by his long-standing desire to solve mathematically interesting physical problems and treat them as if they could be analyzed in principle by a sufficiently powerful computer. There are, to be sure, other relevant factors in this story. Traditions certainly played an important role for instance. But Wilson was not really stuck in any single one and this gave him at best the freedom, not the momentum, to try out new tools and methods. Likewise, the specific local puzzles Wilson worked on certainly led him to explore unconventional field-theoretic routes. But again, these puzzles were largely independent of one another and he could have used different tools for each---as other physicists did at the time---instead of the ones he eventually designed. I will show, in other words, that the key factors in Wilson's part of the story were ``large-scale" pragmatic aspirations and methodological commitments that were sufficiently stable over time to be ultimately effective. 

There will be two additional benefits coming out of this. (i) As we will see, Wilson's aspirations and commitments shaped his prototype in decisive ways. We will thus be able to appreciate more easily what makes his prototype so distinctive. (ii) Some of the ``delicious ironies" filling the early history of effective theories, as Weinberg (\cite*[p. 242]{weinberg_what_1999}) felicitously puts it, will look merely ironic and not incoherent from this perspective. By focusing exclusively on Wilson's concern with understanding the high-energy structure of field theories, for instance, it would be hard to explain why he developed a new kind of field theory, which, as a matter of principle, could not provide any such information. But if we also appeal to Wilson's pragmatic take on physical problems, it becomes easier to understand why he had no qualms about altering the structure of field theory if it meant making some progress.

The paper is organized as follows. The first three sections \ref{sec:The static model: A quick warm-up before the UV}-\ref{sec:From the S-Matrix to model field theories} are devoted to Wilson's trajectory before 1965. Sections \ref{sec:The slicing method in 1965}-\ref{sec:Taking stock: What really made the difference?} analyze Wilson's work on model Hamiltonians in 1965 and its significance. Sections \ref{sec:The road ahead and the meaning of field theory}-\ref{sec:The state of affairs in 1971} briefly examine Wilson's path to his mature conception of effective theories from 1965 to the early 1970s.

\section{The static model: Or how to warm-up before the UV}
\label{sec:The static model: A quick warm-up before the UV}

The story begins in 1956, at Caltech, where Wilson went as a graduate student after finishing his undergraduate studies at Harvard (1954-56). Caltech was an attractive place for aspirant field theorists in the 1950s. Richard Feynman, now a world-renowned physicist after his success with QED, and Murray Gell-Mann, a young rising star who would soon redefine the frontiers of particle physics, had moved there a few years back. The hot topic at the time was to find ways to apply the methods that had worked so well with the electromagnetic interaction to its weak and strong cousins, and both Feynman and Gell-Mann did some work related to this during the 1950s. Unfortunately, Wilson did not interact much with them at the beginning. But his first years at Caltech still gave him the opportunity to embark on the field-theoretic adventure of postwar physics by learning more about nuclear physics and becoming proficient in speaking the language of quantum field theory (QFT).

Wilson started to work on his thesis project in 1958, following Gell-Mann's suggestion to apply the Low equation in the one-meson approximation to K-meson and nucleon scatterings. The main idea was to extend the work that had been done with simplified pion-nucleon models since the early 1950s to the new type of meson, the K-meson or kaon, which had been detected for the first time in cosmic-ray experiments in 1947. Since Wilson did devote most of his thesis to the Low equation---the last part deals with a similar equation but for pion-pion scatterings, it will be useful to provide some background to the key model underlying this equation, the so-called ``static model" or ``fixed-source model", which Wilson kept using up until the early 1970s (see, e.g., \cite{wilson_unpublished_1972}). As he ironically put it years later: 
\begin{quote} 
	Like many second-rate graduate students, I pursued ideas from my thesis topic for over fifteen years before disengaging from it. (\cite[p. 5]{wilson_origins_2005})
\end{quote}
And these ideas, as it turns out, played a crucial role in him developing Nobel-Prize-winning work.

The static model (or, more precisely, the ``static approximation in meson-nucleon models") is part of a large family of relativistic and non-relativistic field-theoretic models in nuclear physics that all trace back to Hideki Yukawa's seminal work in 1935.\footnote{The tradition coming out of Yukawa's work is often referred to as ``meson theory" in the 1950s (e.g., \cite{gell-mann_interactions_1954,wick_introduction_1955}). Major reviews at the time include Pauli (\cite*{pauli_meson_1948}), Blair and Chew (\cite*{blair_subnuclear_1953}), Bethe and de Hoffmann (\cite*{bethe_mesons_1955}), and Wick (\cite*{wick_introduction_1955}). For a short historical account written in the 1950s, see, e.g., Bethe and de Hoffmann (\cite*[chap. 38-9]{bethe_mesons_1955}).} Yukawa's original ambition was to formulate a fundamental theory of the strong interaction by assuming that it is mediated by a new kind of particle, the pions, exactly in the same way as the electromagnetic interaction is mediated by photons (\cite{yukawa_interaction_1935}; see also \cite[esp. chap. 5]{brown_origin_1996}, for more historical details). Over the years, however, this ambition somewhat faded away as physicists faced increasingly serious issues they had not encountered with the electromagnetic interaction: most famously, that the value of the coupling $g_{NN\pi}$ between nucleons and pions in the most realistic relativistic models was seemingly too large to make any sensible use of standard perturbative methods. 

Following Werner Heisenberg, Wolfgang Pauli, and Gregor Wentzel's lead, the main response in the early 1940s was to take the large value of the coupling seriously and use more drastic approximations, such as treating a nucleon as an extended non-relativistic source. This strategy was still suffering from severe issues. The meson field, for instance, was treated in classical terms in the first strong coupling models. After the success of QED in the late 1940s, physicists tried again to formulate relativistic weak coupling models of the strong interaction. But it was quickly found that these models failed to match with significant experiments. The strong coupling models were eventually revived in the early 1950s by Maurice Lévy (\cite*{levy_meson_1952,levy_non-adiabatic_1952,levy_symmetrical_1952}) and others, and physicists started to look seriously at these models again after Geoffrey Chew was able to obtain significant experimental correlations for low-energy pion-nucleon interactions (\cite{chew_comparison_1954,chew_method_1954,chew_renormalization_1954,chew_effective-range_1956,chew_theory_1956}).

The static model is a remarkably simple offspring of this strong coupling tradition (although it may, of course, be used for weak couplings too). The Hamiltonian includes only a simple Yukawa-type interaction term between mesons and a nucleon source, using here the charged scalar static model at the basis of Wilson's 1965 article:
\begin{align}
	H_{\text{static}}= &\frac{1}{(2\pi)^3} \int_{0}^{\infty}  d^3k  \Big[ w_k (a_k^{\dagger}a_k+ b_k^{\dagger}b_k) \Big] \nonumber\\ 
	&+  \frac{g_0}{(2\pi)^3} \int_{0}^{\infty} d^3k \Big[\frac{1}{\sqrt{2w_k}} \big[(a_k+b_k^{\dagger})\tau^{+} + (a_k^{\dagger}+b_k)\tau^{-}\big] \Big],
\label{Eq.1}
\end{align}
where $a_k, a_k^{\dagger}$ (resp. $b_k, b_k^{\dagger}$) are the  $\pi^{+}$-meson (resp. $\pi^{-}$-meson) annihilation and creation operators, $w_k$ the energy of a meson with momentum $k$ and mass $\mu$, $g_0$ the bare Yukawa coupling characterizing the strength of interaction between mesons and the nucleon source, and $\tau^{+/-}$ the bare fixed-source operators accounting for the transition between bare nucleon states $|n\rangle$ and $|p\rangle$ induced by the emission and absorption of mesons. 

As in Yukawa's original proposal, the key idea behind the ``minimal coupling assumption" (in analogy with QED) is to treat the absorption and emission of mesons by the nucleon source one at a time. The simple form of $H_{\text{static}}$ also comes from the ``static approximation": the nucleon source is treated as fixed, i.e., as having an infinite mass. There is thus no need to include kinetic terms for nucleons, and nucleon recoil effects caused by the emission and absorption of mesons can be safely neglected (see, e.g., \cite[pp. 373ff.]{schweber_introduction_1961}, for more details). As Gell-Mann and Watson (\cite*[p. 261]{gell-mann_interactions_1954}) emphasize, this approximation is not indispensable. But it makes the resulting Hamiltonian particularly easy to ``solve", i.e., the energy levels and transition amplitudes between the states of the system are easy to compute, at least approximately since the model specified by Eq. (\ref{Eq.1}) is not exactly solvable. Finally, the model displays ultraviolet (UV) divergences similar to those found in QED, i.e., a direct computation of these energy levels and transition amplitudes involves divergent momentum integrals and thus yields infinite predictions. As we will see, Wilson uses a sharp momentum cut-off in 1965 to make these integrals finite and analyze the model. In the 1950s, by contrast, it was rather common to use a smooth momentum cut-off function $v(k)$ and keep it fixed, which amounts to assuming that the nucleon source is not only static but also spatially extended. 

Chew's main contribution was to bring coupling constant renormalization into the picture (\cite*{chew_pion-nucleon_1953,chew_comparison_1954,chew_method_1954,chew_renormalization_1954}). The older strong coupling models did not involve any, and, as Chew (\cite*[p. 1670]{chew_comparison_1954}) recognized, it did not seem relevant at the time since the cut-off was kept fixed and its value set on the basis of experimental considerations. Chew nonetheless showed that a decisive payoff would come from using the same sorts of renormalization methods that had been used in the case of QED to absorb its UV divergences. He had noticed that the non-relativistic perturbative expansion of matrix elements for pion-nucleon processes contained anomalously large contributions at higher orders. The validity of the perturbative treatment was under threat---higher-order terms were supposed to bring only small corrections to lower-order ones. Chew showed that it was possible to absorb these anomalous contributions in the expression of the propagator and vertex functions, thereby absorbing some of the cut-off dependence of the original model into its parameters (viz. for a restricted set of diagrams and up to a certain order in perturbation theory). He was thus able to re-organize the various terms in the perturbative expansion according to their ``true" relative importance and, along the way, he found some indication that the renormalized coupling $g_r$ would be smaller than the original bare coupling $g_0$ (\cite{chew_method_1954,chew_renormalization_1954}). Since $g_0$ was arbitrary, however, the model had to be confronted with experiments to assess whether the renormalization procedure had improved the original perturbative treatment. Chew (\cite*{chew_comparison_1954}) was able to find significant correlations between the predictions of the renormalized static model and experimental data and thus show that it was more phenomenologically relevant than physicists thought at the time.

Another important idea already apparent in Chew's 1954 articles, and which became even more explicit in his subsequent work with Low, is that the static model and its siblings are unlikely to work at high energies and are best understood as describing only ``low-lying" or low-energy meson states (e.g., \cite[p. 1759]{chew_method_1954}; \cite[p. 1570]{chew_effective-range_1956}). Chew and Low did not simply make these remarks because the static model had a high-energy cut-off built into its structure and excluded key high-energy relativistic effects (e.g., nucleon recoil). The discovery of new strongly interacting particles with a higher rest mass from 1947 onward, such as the aforementioned K-mesons and the hyperons (i.e., heavy baryons), also led to a wider skepticism about the ability of Yukawa's original theory and its descendants to extend to high energies and provide a fundamental description of the strong interaction. As Chew and Low (\cite*[p. 1570]{chew_effective-range_1956}) emphasize, it became natural in this context to use the static model to derive model-independent predictions that might also turn out to be predictions of the true theory (if any). 

The Low equation, sometimes referred to as the Chew-Low equation, is one of the key non-perturbative results that came out of this endeavor (\cite{low_boson-fermion_1955}). This non-linear equation, which is akin to a dispersion relation, relates a given meson-nucleon transition amplitude $A_{if}$ at some energy scale $w$ with all the other amplitudes with the same initial nucleon-meson state $i$ or the same final nucleon-meson state $f$ but with an arbitrary intermediary state of energy $w_k$. There are different versions of this equation, both integral and differential. But schematically, it takes the following form (see, e.g., \cite[sec. 12d]{schweber_introduction_1961}, for more details):
\begin{equation}
A_{if}(w)=\int dw_k \Big(\frac{A_{fk}^{\dagger} A_{ik}}{w_k+w}+ \frac{A_{ik}^{\dagger} A_{fk}}{w_k-w}\Big)
\label{Eq.2}
\end{equation}
By studying its properties, it was then possible to derive general constraints on the structure of meson-nucleon transition amplitudes that would not depend on the specific details of the type of interaction used (such as Gell-Mann and Goldberger's crossing theorem for instance, which relates matrix elements for processes with opposite momenta). And the Low equation could be further simplified by implementing the one-meson approximation, i.e., by neglecting states with multiple mesons (see, e.g., \cite[pp. 401-8]{schweber_introduction_1961}).

So what did Wilson do with the Low equation? On the face of it, he does not appear to have done much apart from showcasing his highly promising skills (as \cite[p. 653]{peskin_ken_2014}, notes too). He mainly studied perturbative expressions of the solutions to the Low equation in the one-meson approximation and showed, in particular, that these expressions take a simplified form at high energies (using different versions of the static model and for some appropriate choice of cut-off when necessary). For instance, in the simplest and somewhat unrealistic case, Wilson found an asymptotic perturbative expression of the form: 
\begin{equation}
	A_{if}(w) \sim \frac{g_s}{1-c g_s \ln(w/\mu)},
	\label{Eq.2b}
\end{equation}
with $g_s$ a power series in the renormalized coupling $g_r$ and $c$ some coefficient (see \cite[part. II, sec. D]{wilson_investigation_1961}, for more details). He also showed that these expressions satisfy key mathematical properties of the exact solutions to the Low equation: for instance, properties of analyticity and unitarity (cf. \cite[pp. 5-6]{wilson_investigation_1961}). And we may ask: if Wilson was so much interested in working in field theory at the beginning of his career as a nuclear physicist, what did really motivate him to pursue this type of dissertation work? After all, as Wilson (\cite*[1/4, 4'47''ff.]{wilson_interview_1991}) recalls, he did most of this work in 1959-60, exactly when both he and his thesis advisor, Gell-Mann, had already lost interest in K-mesons. And, strictly speaking, Wilson did not really attempt to extend the Low equation to the particular case of K-mesons, as Gell-Mann had originally suggested. The main concrete puzzle he was working on at the time was the issue---rather unappealing for an aspirant field theorist---of analyzing the mathematical properties of the solutions to a phenomenological equation with the help of perturbative methods.

The project Wilson devised for himself was, in fact, more ambitious. He wanted to use the static model and the equations derived from it to understand the general structure of realistic field theories and their solutions. Wilson is explicit about it both in his dissertation and in his correspondence with Gell-Mann (e.g., \cite[p. 4]{wilson_letter_1959-1}; \cite*[pp. 14, 16]{wilson_investigation_1961}). In the introduction of his dissertation, for instance, he emphasizes that studying the convergence and analytic properties of the perturbative solutions to the Low equation in the one-meson approximation could ``easily be relevant to the far more difficult problem of covariant field theory" (\cite[p. 16]{wilson_investigation_1961}). 

The idea was indeed not entirely out of place. As Gian-Carlo Wick (\cite*[p. 339]{wick_introduction_1955}) had already emphasized, the static model had a ``fighting chance" compared to other models of the strong interaction: at least some versions of the model had appropriate quantum numbers to account for existing particles and had made clear contact with experiments. The static model was also displaying similar UV divergences and was amenable to similar perturbative renormalization methods as realistic theories. The perturbative expressions derived from the static model at high energies were even similar to the type of perturbative series that had been found in QED, say, $a_1g+a_2g^2\ln(w)+a_3g^3\ln^2(w)+...$, with $w$ some energy scale, $g$ some coupling, and $a_i$ some coefficients (cf. \cite[p. 17]{wilson_investigation_1961}). So if the static model was solved by using improved perturbative methods, or even non-perturbative ones, it was reasonable to expect that these methods would also work for realistic theories. Finally, the minimal coupling and static approximations made the static model easy to compute with. As Francis Halpern and his collaborators (\cite*[p. 155]{halpern_physical_1959}) emphasize, this was, in fact, one of the main motivations for studying the static model in the late 1950s, before physicists started to move away from field-theoretic treatments of the strong interaction. And all these similarities and the easiness with which one could work with the static model convinced Wilson that it would be a fruitful place to learn more about the high-energy structure of QFT.\footnote{One might be worried here that Wilson would not be able to learn anything about the high-energy structure of QFT from the static model insofar as it was merely a \textit{low-energy} model, as Chew had already emphasized. I will discuss further this point below to support the idea that Wilson's main goal in 1965 was rather to develop a new non-perturbative method than to understand directly the high-energy structure of QFT. But the same argument could already be made here, in 1959-60. As it turns out, Wilson does acknowledge later on the discrepancy between his main field-theoretic aspiration and the tools he was using to fulfill it at the time: ``Murray's goal was to use the equation to help make sense of the phenomenology of K-p scattering. But I became fascinated with the high-energy behavior of solutions to the Low equation, despite its being a reasonable approximation for physics, if at all, \textit{only for low energies}." (\cite[p. 5]{wilson_origins_2005}, my emphasis; see also \cite*[p. 589]{wilson_renormalization_1983})}

\section{The impact of Gell-Mann and Low's early renormalization group picture}
\label{sec:The impact of Gell-Mann and Low's early renormalization group picture}

Wilson's simplified perturbative expressions at high energies were, in particular, remarkably similar to several results that Gell-Mann and Low had found in the case of QED, and this led Wilson to become increasingly interested in their work. 

As we have already seen, Gell-Mann and Low were both working on meson-nucleon models in the mid-1950s (e.g., \cite{gell-mann_interactions_1954,low_boson-fermion_1955}). But around the same time, they also wrote an article investigating the high-energy behavior of QED, which could, or so they thought, be relevant to meson theory (\cite[p. 1301]{gell-mann_quantum_1954}). QFTs had been known to display UV divergences since the late 1920s. The perturbative renormalization methods developed in the late 1940s for QED had brought some relief. But besides being mathematically and conceptually dubious, it was not clear whether these methods were not hiding the real problem, to wit, the seemingly pathological high-energy behavior of QFTs, and Gell-Mann and Low wanted to gain some insight into it. Their article has since become known as one of the first instances in which RG methods were introduced. It is worth emphasizing, however, that Gell-Mann and Low were not concerned with, or even talking about, the RG at this point (as it is explicit from the title and content of the article and as Gell-Mann also emphasizes later on in \cite*[p. 276]{gell-mann_particle_2010}). They were interested in developing a new method to understand the high-energy behavior of the renormalized photon and electron propagators (obtained with the standard renormalization methods used at the time).

Gell-Mann and Low were confronted with two immediate issues. First, the propagators were displaying infrared singularities when the mass $m$ of the electron was taken to zero. For instance, the first few lowest-order terms for the renormalized electron propagator included contributions depending on $\ln(k/m)$, with $k$ the four-momentum of an electron, and one could not naively take $m \to 0$ to derive a simplified asymptotic expression for $k \gg m$. Second, the increasingly large contributions of higher-order terms at high energies prevented one from identifying a subset of dominant terms and obtaining a simple asymptotic expression from the perturbative series. At the same time, it was also clear to Gell-Mann and Low that logarithmic radiative corrections in QED would bring deviations from the naive high-energy scaling behavior expected from the expression of the bare propagators (e.g., in $1/k$ for electrons). The underlying problem, then, was to find the general dependence of renormalized quantities on the momentum and mass scales involved in the process of interest (i.e., on $k/m$ for the electron propagator).

Gell-Mann and Low's strategy to solve this problem is easier to appreciate in the simple case of the electron propagator without photon self-energy contributions (cf. \cite[sec. 3]{gell-mann_quantum_1954}). First, a direct calculation shows that the bare electron propagator $s(\lambda, k, m)$ with a cut-off $\lambda$ is not singular for $m \to 0$. By contrast, the renormalized propagator $s_r(k, m)$ and the wave function renormalization factor $z(\lambda, m)$ both take the form of a perturbative series including singular contributions depending on $\ln(k/m)$ and $\ln(\lambda/m)$, respectively. It should thus be possible to express $s_r$ and $z$ in such a way that these contributions directly cancel each other in the equation $s(\lambda, k, m)= z(\lambda, m) s_r(k, m)$.

Then, one takes the asymptotic limit $\lambda \gg k \gg m$ in the expression of $s(\lambda, k, m)$ to remove any non-singular mass dependence and replace $s(\lambda, k, m)$ with its simplified asymptotic expression in the previous equation (after removing factors in $1/k$ on both sides). The result is a general asymptotic functional equation for $s_r(k/m)$ of the form $s(\lambda/k)= z(\lambda/m) s_r(k/m)$. 

Finally, one can derive the general form of the solutions to this equation and obtain information about the asymptotic behavior of $s_r(k/m)$ for $k \gg m$. If photon self-energy contributions are included and the dependence on the renormalized coupling $e_r$ is made explicit, the solutions take the form:
\begin{equation}
	s_r(k/m, e_r)=A(e_r) H\big[\frac{k}{m} \phi (e_r)\big],
	\label{Eq.2c}
\end{equation}
with $A$, $H$ and $\phi$ some unknown functions. Similar expressions can be found for the photon propagator and the coupling (cf. \cite[p. 1307]{gell-mann_quantum_1954}). And by analyzing how the unknown functions entering into these expressions vary with $e_r$ and $k$, say, by computing their expression up to the first or second order in perturbation theory, it is possible to draw more perspicuous conclusions about the asymptotic behavior of renormalized quantities in QED than allowed by their original perturbative expression (see \cite[sec. 5]{gell-mann_quantum_1954}, for more details).

As Wilson (\cite*[p. 1843]{wilson_renormalization_1971-1}) notes later on, Gell-Mann and Low's method is somewhat convoluted. In particular, their most important result by today's lights, namely, an ``RG equation" governing the variation of couplings across scales of the form $\lambda de(\lambda)/d\lambda=\beta(e(\lambda))$ for some function $\beta$ and some momentum parameter $\lambda$, is independent of the original issue of infrared singularities. Still, Wilson realized that the simplified asymptotic expressions he had derived in his thesis were displaying the same kind of scale and coupling dependence. For instance, if one substitutes $k/m$ for $w/\mu$ and $e_r$ for $g_r$ and assumes that the power series $g_s$ is an analytic function $F$ of the renormalized coupling $g_r$, one finds that Eq. (\ref{Eq.2b}) takes the same form as Eq. (\ref{Eq.2c}) with:
\begin{align}
	A(e_r) &= -1/c \nonumber \\
	H(\frac{k}{m},e_r) &= \Big[\ln(\frac{k}{m} \phi (e_r))\Big]^{-1} \nonumber\\
	\phi(e_r) &= \exp\Big[-\big(cF(e_r)\big)^{-1}\Big]
	\label{Eq.2d}
\end{align}
This is, of course, a somewhat oversimplified reconstruction. Wilson provides more details in his dissertation (see \cite*[part II, sec. F]{wilson_investigation_1961}) and summarizes the connection as follows later on: 
\begin{quote} 
	Well at any rate, I had run through the series to enough orders to see that there was a rule there. You could complete the rule and then you could complete the sum for the high-energy limit of the solution of some of these equations. And that sort of got me interested in Gell-Mann-Low. (\cite[1/4, 7'34''ff.]{wilson_interview_1991})
\end{quote}

These similarities certainly explain why Wilson became interested in Gell-Mann and Low's article. But we might still wonder whether it had any real impact on his early works. Wilson has repeatedly emphasized in later interviews that his study of the asymptotic properties of the solutions to the Low equation was his ``first introduction" to the RG (e.g., \cite[p. 18]{lubkin_wilson_1982}; \cite[p. 572]{lewin_nobel_1995}; \cite[part I]{wilson_interview_2002}). He even emphasized in a letter to Wolfhart Zimmermann that Gell-Mann and Low's article was one of his main sources of inspiration for the work he did in 1965 (\cite{wilson_letter_1965}; see also \cite*[p. 589]{wilson_renormalization_1983}). Yet it does not seem that Gell-Mann and Low's article played such a crucial role. First, Wilson briefly showed in his dissertation how their ideas could be applied to the static model by lumping appropriate pieces of matrix elements together (cf. \cite*[pp. 43-8]{wilson_investigation_1961}). His treatment, however, was rather peripheral and the method he used was also quite different from what he did in 1965 (more details will be given in section \ref{sec:The slicing method in 1965}). For instance, Wilson was far from introducing large energy gaps in the state-space of the static model at this point. Second, Wilson understood Gell-Mann and Low's ideas in formal terms at that time compared to his own understanding of the RG after 1965. According to Wilson, for instance, one of their key moves was to introduce an ``alternative unit of length" $\lambda$ to renormalize quantities in QED and safely take the zero-mass limit (\cite*[p. 43]{wilson_investigation_1961}). But he did not yet associate any physical picture with their new arbitrary ``renormalization condition", say, as resulting from a coarse-graining procedure.\footnote{Note that there is some ambiguity about how to best interpret the two ``cut-off" parameters $\lambda$ and $\lambda'$ in section 4 of Gell-Mann and Low's article. I am inclined to interpret them as arbitrary subtraction points given how Gell-Mann and Low introduce and use them (cf. \cite[pp. 1305-6]{gell-mann_quantum_1954}). This also seems to be Wilson's interpretation in his dissertation (\cite*[p. 43]{wilson_investigation_1961}).} 

Suppose that Wilson did get something both concrete and substantial out of Gell-Mann and Low's article. We might still wonder whether it was in any way faithful to what they had achieved at the time by today's lights. Gell-Mann and Low's article is notoriously difficult to understand and the short account I provided above only scratches the surface. In hindsight, they are often seen as having shown that the structure of renormalizable theories is somewhat arbitrary (e.g., \cite[chap. 23]{schwartz_quantum_2013}).\footnote{See Wilson (\cite*{wilson_model_1970,wilson_renormalization_1971-1}) and Weinberg (\cite*{weinberg_why_1983}) for their own respective exegesis, and Fraser (\cite*{fraser_twin_2021}) for a more comprehensive historical account. See also Rivat (\cite*{rivat_renormalization_2019}) for more details about the difference between Wilson's and Gell-Mann and Low's RG in contemporary physics, and Williams (\cite*[p. 7]{williams_historical_2016}) for a complementary reading of Gell-Mann and Low's achievements in 1954 emphasizing ideas of scale invariance and broken scale invariance.} One can use different renormalization conditions, each parametrized by a different value of some arbitrary momentum scale, and adjust the structure of the renormalized theory at stake accordingly. One can also rely on this arbitrariness to derive RG equations for renormalized couplings, probe their high-energy behavior, and show that this behavior is independent of the initial bare parametrization of the theory. Finally, one can eliminate large contributions that spoil the perturbative treatment of the theory, the so-called ``large logarithms problem", by imposing a suitable renormalization condition.

Now, whether these hidden gems were already clear in Gell-Mann and Low's mind in 1954, there are good reasons to think that Wilson did not fully grasp them, at least not before 1965. First, as Wilson acknowledges later on, he could not ``understand the paper of Gell-Mann and Low" and had to ``reinvent the renormalization group before [he] could understand [it]" (\cite[1/4, 5'34''ff., 26'44''ff.]{wilson_interview_1991}). To be sure, Wilson's memory and later remarks should be treated with caution, and especially this one, as he certainly did ``understand" enough of Gell-Mann and Low's work to be able to apply it in his dissertation. However, he may not have understood what was the central point of their article. For instance, Wilson was much more aware in 1971 than he originally was in 1961 that the issue of infrared singularities is irrelevant for assessing the scaling behavior of quantities in QED (e.g., compare \cite[pp. 43-8]{wilson_investigation_1961}, and \cite*[p. 1843]{wilson_renormalization_1971-1}). Second, it is not as if Wilson could easily get these hidden gems from Gell-Mann himself. Wilson did most of the work that was closely related to Gell-Mann and Low's RG in 1959-60, when Gell-Mann was visiting Lévy in Paris and Wilson entered the Harvard Society of Fellows (1959-62). And it does not seem that Wilson ever exchanged on the topic of the RG with Low too (\cite[1/4, 8'09''ff.]{wilson_interview_1991}). In a word: Wilson probably had to work out Gell-Mann and Low's article on his own and extract from it whatever could be relevant for his own work.\footnote{There are also uncertainties concerning the impact of Bogoliubov and Shirkov's introduction to QFT on Wilson's work, which was the first textbook to include a section on the RG (\cite[chap. 9]{bogoliubov_introduction_1959}) and which Wilson cites in his dissertation (\cite*[p. 46]{wilson_investigation_1961}). The book was first published in English in 1959 and copies were apparently circulating beforehand. Wilson sometimes mentions that this book was already used during his coursework and, most crucially, before he realized the connection between his dissertation work and Gell-Mann and Low's article (\cite[1/4, 8'00''ff.]{wilson_interview_1991}). Sometimes he emphasizes that he read the book on his own and that it was not covered during his coursework (\cite[Part I]{wilson_interview_2002}). To make the matter even worse, Wilson points out that he did not ``understand" Bogoliubov and Shirkov's introduction to the RG either (\cite[1/4, 5'45''ff.]{wilson_interview_1991}).}

The most plausible reading, in my sense, is that Wilson was influenced by Gell-Mann and Low's article in two ways before 1965. First, Wilson was driven to think more deeply about the need to develop new methods to investigate the high-energy structure of QFTs. After all, this was one of Gell-Mann and Low's most explicit goals in the article---Wilson could not have missed it. They were also explicit about the academic character of their investigation for the specific case of QED given its unlikely empirical validity at high energies (\cite[p. 1301]{gell-mann_quantum_1954}). That is: the underlying goal of the article was to develop a new method for a future fundamental QFT and not to understand the high-energy structure of QED \textit{per se}. Second, Gell-Mann and Low's article led Wilson to think about the idea of a \textit{sequence} of renormalized couplings associated with different scales as he explicitly acknowledges in the draft of his 1965 article (\cite[p. 2, footnote 1]{wilson_model_1965-1}). Physicists were using different types of renormalization conditions at the time (see, e.g., \cite[part IIC]{schweber_mesons_1955}). Contrary to Gell-Mann and Low's approach, however, all these conditions were $\textit{fixed}$, i.e., specified at a particular scale. That said, it is not as if Wilson took his scale-dependent physical picture of the RG, already explicit in 1965 as we will see, from Gell-Mann and Low's work. They only gave him a ``very strong direction" for his investigations, as he puts it later on (\cite*[2/4, 32'23''ff.]{wilson_interview_1991}), but not an explicit template with a fully-fledged physical interpretation.

\section{From the S-matrix to model field theories}
\label{sec:From the S-Matrix to model field theories}

Wilson eventually defended his dissertation in June 1960 while he was a Junior Fellow at Harvard. The years 1959-63 were the occasion for him to travel to different places and interact with various theoretical and experimental physicists. He had visited Berkeley and Stanford in 1959 to discuss his dissertation work with Geoffrey Chew, Stanley Mandelstam, and Marshall Baker. His time at Harvard also gave him the opportunity to interact with Kenneth Johnson and Francis Low at MIT. He spent as well the year 1962 at CERN, partly as a Ford Foundation Fellow, and finally moved to Cornell as a junior faculty in 1963.

Wilson worked mainly on topics related to the rising S-matrix program during those years. By 1963, however, he had convinced himself that this work would lead him nowhere. He decided to return to field theory and devote his time to understanding the high-energy structure of QFT in the context of the strong interaction. As Wilson recounts in his Nobel Prize lecture:
\begin{quote} 
	I rejected S-matrix theory because the equations of S-matrix theory, even if one could write them down, were too complicated and inelegant to be a theory; in contrast, the existence of a strong coupling approximation as well as a weak coupling approximation to fixed source meson theory helped me believe that quantum field theory might make sense. (\cite*[p. 590]{wilson_renormalization_1983}; see also \cite*[1/4, 13'23''ff.]{wilson_interview_1991})
\end{quote}
It is important to note here that Wilson's decision was not merely the result of his ultimate distaste for the S-matrix program and the encouraging signs that had emerged from his dissertation work. He had also been driven toward physics since high school by his desire to work on ``interesting and productive mathematical problems" and he saw the high-energy behavior of QFT as the ``most interesting and useful" one he could work on at the time (\cite[p. 5]{wilson_origins_2005}). But this decision, though judicious as it may look with the benefit of hindsight, was nonetheless going to make him somewhat isolated from the most popular areas of research in nuclear physics during this period, the S-matrix and current algebra programs.

There was indeed a growing skepticism toward the ability of QFT to provide a coherent and fundamental description of particles in the early 1960s. The 1950s had seen heated debates as to whether QED remains consistent across all scales (see \cite{blum_decline_2021}, for more details). Many influential physicists, including Wolfgang Pauli and Lev Landau, had provided arguments to the effect that it would not. For instance, in today's terms, Landau showed that the perturbatively renormalized expression of the coupling in QED would become infinite at some finite high-energy scale. And since QED was supposed to be the field-theoretic blueprint of particle physics, these physicists had considered that the same situation would obtain for the weak and the strong interaction, if they were ever going to be described by fundamental QFTs. Similar arguments were put forward in the context of the strong interaction with toy model field theories, such as the Lee model (\cite*{lee_special_1954}), which led Landau to declare:  
\begin{quote} We are driven to the conclusion that the Hamiltonian method for strong interaction is dead and must be buried, although of course with deserved honor. (\cite*[p. 801]{landau_fundamental_1960})
\end{quote}
Meson theory was also becoming less and less popular. Even Chew, who had been one of the most active physicists working in this area, had somewhat lost interest in the static model by the end of the 1950s, as Wilson had the misfortune to discover when he visited Berkeley (cf. \cite*{wilson_letter_1959}).\footnote{To be fair, there were still many physicists working within the traditional field-theoretic framework in the early 1960s (e.g., \cite{salam_weak_1959,glashow_partial-symmetries_1961,schwinger_field_1964}). There were also ``public encouragements" by influential physicists to do so. For instance, as Victor Weisskopf put it in his concluding remarks to the 11th Rochester conference held at CERN in July 1962: ``In fact, it could be that all that we have found follows from field theory. And this is why one should encourage those people who are very faithful to field theory and go on with a painstaking study of field theory in all its aspects. Perhaps they make little progress, but what they find might be all the more significant" (\cite*[p. 933]{weisskopf_concluding_1962}). Since Wilson was at CERN at the time, there is actually some chance that he heard these rather prophetic words, or at least read them in the proceedings afterward. Wilson did attend the conference (as it would be common for visiting scientists) and was, in particular, secretary for the session ``Regge poles and related topics" (cf. \cite[p. 501]{prentki_proceedings_1962}). Wilson (\cite*[1/4, 20'53''ff.]{wilson_interview_1991}) also explicitly refers to this specific conference in his interview with Cao.}

In this context, toy model field theories provided a natural framework to investigate further the high-energy behavior of QFTs (also referred to as ``model field theories" in, e.g., \cite{bludman_unified_1966}). Following Tsung-Dao Lee's work, Walter Thirring (\cite*{thirring_soluble_1958}) had proposed another exactly solvable simple model. Johnson (\cite*{johnson_solution_1961}) had further clarified the structure of its correlation functions. And, as Wilson suggests in his interview with the Physics of Scale project in 2002, he got increasingly interested in the structure of these model field theories thanks to his interaction with Johnson and other physicists at MIT in the early 1960s.\footnote{``There was hardly anybody at Harvard. So what I used to do was go down and eat lunch with people at MIT: Francis Low, Ken Johnson and those people. And that was very interesting. That set me up for what got me going in early 1963/1964. Partly it was looking at the work of Johnson and Marshall Baker and trying to figure out why I didn't agree with it" (\cite[part I]{wilson_interview_2002}). See also Huang (\cite*[p. 20]{huang_critical_2015}).} Wilson eventually started to work ``in earnest" on the high-energy behavior of QFTs during the summer 1963 while he was ``sitting in a hospital in Yugoslavia" trying to recover from ``the usual Third World stomach problems" (\cite*[1/4, 15'00''ff.]{wilson_interview_1991}). He was drawn on this occasion to look at the Lee model, a partially relativistic pion-nucleon model very much akin to the static model and which he eventually put to use in 1965. It had already been known at least since the mid-1950s that the Lee model was not even close to being realistic (e.g., \cite[p. 339]{wick_introduction_1955}). Yet, this model was still sufficiently rich in non-trivial properties, such as UV divergences and coupling constant renormalization, to appear relevant for analyzing the mathematical structure of QFT (\cite[p. 1329]{lee_special_1954}; \cite[p. 880]{ruijgrok_exactly_1956}). And the fact that it was exactly solvable made it extremely attractive for identifying perturbative artifacts within the set of issues plaguing QFTs.

Lee, Thirring, and Johnson were, of course, not the only ones trying to clarify the mathematical structure of QFTs at the time. The axiomatic field-theoretic tradition, which Wilson became acquainted with in the early 1960s, had been burgeoning since the mid-1950s thanks in particular to Arthur Wightman's and Rudolf Haag's works. And there is another key player closely related to this tradition in Wilson's story, namely, Léon van Hove. He was one of the few mathematically-oriented field theorists who had relied on the static model for foundational work. Even more striking is the fact that he had justified using this model on very similar grounds as Lee (\cite*[p. 1329]{lee_special_1954}) and Wilson (\cite*{wilson_model_1965}) did afterward:  
\begin{quote} 
	This method [i.e., perturbative renormalization], which leaves divergences in the fundamental constants of the theory, cannot be considered as definitive, and the search for a more exact method, as well as its comparison with the method of perturbative renormalization, appear to be indispensable tasks. An interesting direction of research regarding the difficulties arising in this method is given by the study of certain field models, simple enough to be susceptible to rigorous treatment, but which nevertheless present in an attenuated form the divergences that are characteristic of general cases. A well-known example is provided by a neutral scalar field (boson field) in scalar interaction with point sources too heavy to move during the emission or absorption of bosons [i.e., the static model].\footnote{``Cette méthode, qui laisse subsister les divergences dans les constantes fondamentales de la théorie, ne peut être regardée comme définitive, et la recherche d'une méthode plus exacte, ainsi que sa comparaison avec la méthode de perturbation et renormalisation, semblent des tâches indispensables. Une orientation intéressante concernant les difficultés qui s'y présentent est fournie par l'étude de certains modèles de champs, assez simples pour être susceptibles d'un traitement rigoureux, mais présentant néanmoins sous une forme atténuée les divergences caractéristiques des cas généraux. Un exemple bien connu en est fourni par un champ scalaire neutre (champ de boson) en interaction scalaire avec des sources ponctuelles trop lourdes pour se mouvoir lors de l'émission ou de l'absorption des bosons."} (\cite[p. 146]{van_hove_les_1952}, my translation)
\end{quote}
As it turns out, van Hove was the head of the Theory Division at CERN during Wilson's stay in 1962. They certainly did interact at the time.\footnote{Wilson thanks van Hove and his group for their hospitality in (\cite*[p. 44]{wilson_regge_1963}) and briefly mentions later on that he had been ``in touch with everybody" in the group during his visit (\cite*[1/4, 14'51''ff.]{wilson_interview_1991}).} And there is also clear evidence from a correspondence between Wilson and Zimmermann that Wilson read van Hove's 1952 article before working on model Hamiltonians in 1965 (cf. \cite{wilson_letter_1965}).

So what was the impact of this model field-theoretic tradition on Wilson's works? Looking ahead, Wilson came around this time to build for himself a portfolio of toy model field theories to investigate the structure of QFTs, including the charged scalar static model, the Lee model, and the $\phi^4$-theory (cf. \cite{wilson_construction_1965,wilson_model_1965-1,wilson_model_1965}). None of these models would bring any definitive conclusion by itself, say, about the ultimate consistency of QFT or the relevance of a new non-perturbative method. However, they would constitute a ``theorist's laboratory" in which to perform various ``experiments" (\cite[p. 6]{wilson_origins_2005}). Wilson could then examine how a particular model would respond, as it were, to a new method, or how its perturbative solutions would behave at high energies, and convince himself, if he found similarities across models, that the theoretical phenomenon at stake was robust.

Now, as Wilson recognizes in his letter to Zimmerman, van Hove's 1952 article had a decisive impact on his attempt to ``construct a relatively simple intuitive picture of the structure of quantum field theory" in his 1965 article. Van Hove's primary concern was a result akin to what is now known as ``Haag's theorem": namely, that the systems described by the free and interacting versions of a QFT model with an infinite number of degrees of freedom have unitarily inequivalent state-spaces, which raises the issue of whether the solutions obtained by applying perturbation theory to the free model have anything to do with the exact ones. Van Hove knew that this result would not hold in the case of a model with only a finite number of degrees of freedom and this led him to suggest a simple method for evaluating the validity of perturbation theory in the corresponding infinite model (\cite*[sec. 5, esp. p. 155]{van_hove_les_1952}). One would first need to compute the quantity of interest $Q$ in the finite model with a high-energy cut-off $\Lambda$ up to the order $n$ in the coupling $g$: for instance, 
\begin{equation}
	Q(\Lambda, n)=a_1g+a_2g^2\ln(\Lambda)+ ... + a_ng^n\ln^{n-1}(\Lambda),
	\label{Eq.2e}
\end{equation}
using here the same notation as before. One would then define some maximal bound $G$ for the coupling, i.e., $ 0 < g < G$, under which the error $\epsilon$ resulting from neglecting the perturbative term in $g^{n+1}$ would not be too large, i.e., $\epsilon \leq \epsilon_{\text{max}}$. In the simple example above, $G$ would be implicitly defined by $\epsilon_{\text{max}}= a_{n+1}G^{n+1}\ln^{n}(\Lambda)$. If one found $G(\Lambda,\epsilon_{\text{max}}, n) \to 0$ in the infinite cut-off limit $\Lambda \to \infty$, the perturbative treatment of the quantity would be invalid.

As we will see in more detail in section \ref{sec:The slicing method in 1965}, Wilson's approach in 1965 was quite different from what van Hove had suggested. First, Wilson did not take the exact solvability of a model to be significant (\cite*[p. 162]{wilson_construction_1965}). There were, in Wilson's eyes, too few models known to be exactly solvable at the time and none of them were realistic. Overall, Wilson was more interested in developing methods for constructing and solving a model than finding the solutions of the model themselves (cf. \cite{wilson_construction_1965,wilson_model_1965-1,wilson_model_1965}). For van Hove, by contrast, the exact solvability of a model was essential for assessing the deficiencies of perturbation theory---hence the choice of a simple and exactly solvable static model with only one neutral scalar field and a large-distance cut-off. Second, Wilson reduced the state-space of his version of the static model to a set of well-separated continuous slices in 1965 while van Hove had suggested using a model with a finite state-space, which amounts, in today's terms, to working with a lattice of finite extent. It was common to impose both a large-distance and a short-distance cut-off in the 1950-60s (e.g., \cite[pp. 340-2]{wick_introduction_1955}). But it was mostly a matter of convenience and the perturbative quantities of interest, such as the matrix elements for some scattering process, were calculated by integrating over all possible momenta (e.g., \cite[sec. 3-6]{wick_introduction_1955}; \cite[part III]{henley_elementary_1962}). In van Hove's work, by contrast, the difference between a finite and an infinite number of degrees of freedom had acquired a foundational significance. Finally, Wilson's method in 1965 involved evaluating how low-energy degrees of freedom affect the lowest energy levels of a Hamiltonian describing a physical system at some fixed high-energy scale while van Hove's suggestion amounted to evaluating whether higher-order terms in perturbation theory become too large at high energies.

Having said that, van Hove's work still influenced Wilson in decisive ways. Wilson is not explicit about its specific impact in his letter to Zimmerman. But the central result of van Hove's 1952 article certainly led Wilson to realize that perturbation theory was even more ill-behaved in the case of infinite models than he had originally thought. He must have felt even more keenly the need to develop new methods to go beyond standard applications of perturbation theory. In the same vein, van Hove's work must have had a liberating effect on Wilson's mind. The use of non-relativistic finite models to analyze field-theoretic issues was indeed not very fashionable in those days (see, e.g., George Sudarshan's and Werner Heisenberg's skeptical reaction to Wilson's talk on model Hamiltonians in \cite*[pp. 172-4]{wilson_construction_1965}). Van Hove's method and suggestions must have given Wilson both a clear precedent and an incentive for cutting in the way he deemed necessary the infinite-dimensional state-space of the static model. But most importantly, van Hove made the key suggestion that the perturbative results obtained from the infinite static model could be evaluated by relying on a sequence of finite versions of the model, each including new high-energy degrees of freedom, and assessing whether the perturbative treatment was still reliable in the infinite cut-off limit. Wilson did frame the issue of assessing whether the solutions of a model exist in those terms in the talk he gave in 1965 (cf. \cite[pp. 161-2]{wilson_construction_1965}). And, as we will see in the next section, Wilson's strategy in his 1965 article does bear the mark of van Hove's suggestion. 

We have at this point all the main elements that seem to be indispensable to understanding Wilson's path to his 1965 article. In particular, Wilson wants to work on interesting mathematical problems related to the high-energy structure of field theory. He has been fiddling with the static model for many years now. He knows all too well the need to invent new methods, as Gell-Mann and Low did, to go beyond standard applications of perturbation theory. He has also stored various toy models in his theoretical laboratory along the way. And since he is aware that they either have no practical relevance or are unlikely to remain empirically valid across all scales, there is no reason left for him not to try to slice them up in novel ways and examine their behavior at high energies.\footnote{I will leave aside Wilson's work on the operator product expansion (OPE) in 1963-64 insofar as it does not seem to have influenced much his work on model Hamiltonians in 1964-1965. See, e.g., Wilson (\cite*[p. 26]{wilson_cornell_1971}; \cite*[part II]{wilson_interview_2002}) for some textual evidence supporting the idea that he was indeed working on two distinct attempts to probe the structure of QFTs in 1963-65.}

\section{The slicing method in 1965}
\label{sec:The slicing method in 1965}

\begin{quote} 
	We have not used perturbation theory---we have used an axe on the Hamiltonian. (Wilson, ``What Every Quantum Mechanic Should Know")
\end{quote}
 
\noindent Wilson made this remark to his students in the early 1980s while teaching quantum mechanics.\footnote{Cf. Kenneth G. Wilson Papers, \#14-22-4086. Division of Rare and Manuscript Collections, Cornell University Library. Box 1, Folder 12.} Yet it illustrates well what must have been his state of mind in June 1965, when he first laid down his early formulation of the RG and EFTs in publishable form (\cite{wilson_model_1965}). Wilson was indeed explicit about his main goal. He meant to develop a new non-perturbative method to extract qualitative information from realistic field-theoretic models of the strong interaction. Standard perturbative methods had been shown to be inapplicable. The usual rules for assessing whether other existing approximation methods, such as the variational method and the WKB approximation, would be appropriate were not working too. And following Lee's work, it was not even clear whether these models would ultimately make any sense (\cite[p. B446]{wilson_model_1965}). New methods were thus required to understand their structure and more generally the structure of field theory itself---not perturbative field theory. In hindsight, Wilson's achievements are remarkable and it will be useful to retrace his steps in some detail.

Wilson starts by laying down a version of the static model with two meson fields of opposite charge interacting with a two-state nucleon source (cf. Eq. \ref{Eq.1}). These models are old hat in 1965. But as we have seen, the model field-theoretic tradition has given them some respectability for investigating the mathematical structure of QFT. Accordingly, the most important feature of Wilson's model Hamiltonian is that it displays the same sorts of issues as realistic QFTs. But its simplicity, its ability to encode all the relevant properties of the target system, its intuitive physical character, and the existence of many approximation methods to work out its consequences also make it a very appealing tool in Wilson's eyes, much more than the S-matrix machinery for instance (\cite[pp. B446-7]{wilson_model_1965}).

Wilson's first key move is to reduce the state-space of the model $H_{\text{static}}$ in Eq. (\ref{Eq.1}) to a set of well-separated continuous ``slices" parametrized by some large scale $\Lambda$ (\cite*[p. B447]{wilson_model_1965}):
\begin{equation}
	0 < k < k_0 \; , \; \frac{1}{2} \Lambda < k < \Lambda \; , \;  ... \; , \;  \frac{1}{2} \Lambda^n < k < \Lambda^n \; , \;  ... \; ,
\label{Eq.3}
\end{equation}
where $k$ stands for the possible momenta of mesons and $k_0$ for some momentum scale of the order of the meson mass $\mu$ and much smaller than $\Lambda$, using here the natural unit $\mu=1$ for simplicity (i.e., $\Lambda^n \equiv \mu \Lambda^n$). The large scale-separation between the slices ensures that low-energy mesons introduce only a small shift in the energy levels of the system composed of the static source and high-energy mesons. The system at the scale $\Lambda^n$ can thus be studied by treating the effects of mesons with characteristic scale $\Lambda^{n-1}$, ..., $k_0$ as a series of incredibly refined perturbations in powers of $1/\Lambda$ (\cite[p. B447]{wilson_model_1965}).

By performing this first step, Wilson has also reduced the general problem of solving $H_{\text{static}}$ into two main tasks: (i) solve the partial Hamiltonian in each slice; (ii) evaluate how they relate to one another. The system composed of the source and low-energy ``laboratory" mesons of momenta $k \in ]0, k_0[$ is the only one that does not receive small perturbations from the other slices. It thus appears natural that Wilson would start by solving the Hamiltonian $H_{\text{lab}}$ associated with this system. And since $H_{\text{lab}}$ has a fixed cut-off $k_0$, the traditional tools of quantum mechanics can be used to obtain approximate solutions. The choice of approximation method depends only on the value of $g_0$ in Eq. (\ref{Eq.1}). 

To connect the discussion with standard perturbative methods, let us assume that the coupling $g_0$ is small enough to allow for a perturbative treatment in the coupling constant. The first thing to do is to find the eigenstates and eigenvalues of the free part $H_{\text{free}}$ of $H_{\text{lab}}$ (e.g., plane wave solutions with energy $w_k$). The application of $a_k^{\dagger}$ and $b_k^{\dagger}$ to the ground state $|0\rangle$, i.e., the bare nucleon state with no mesons, generates a complete basis $\{|\phi_n \rangle\}=\{|0 \rangle, |k_1 \rangle, |k_1, k_2\rangle, \text{etc.}\}$, where $|k_1, ..., k_m\rangle$ stands for the state of the nucleon source with $m$ mesons of momenta $k_i$. For simplicity, I ignore the difference between charged mesons and the particular state of the fixed source here. Now suppose that the interaction term $g_0V$ in the full Hamiltonian $H_{\text{lab}}=H_{\text{free}}+g_0V$ brings only small deviations to the free solutions, i.e., that the eigenstate $|\psi\rangle$ of $H_{\text{lab}}$ with energy $E$ reduces to a free state $|\phi\rangle$ of energy $w$ in the limit $g_0 \to 0$. Then, we can use standard non-relativistic methods to derive approximate expressions for $|\psi\rangle$ and $E$ at any order in perturbation theory. For instance, in the case of a state $|\psi\rangle$ associated with a free one-meson particle state $|k \rangle$, $E$ takes the following schematic form: 
\begin{align}
	E= w_k + & g_0V_{k k}+  g_0^2\int_{k'} \frac{V_{k k'} V_{k' k}}{w_k-w_{k'}} \nonumber \\
	&+ g_0^3\int_{k' k''} \frac{V_{k k'} V_{k' k''}V_{k'' k}}{(w_k-w_{k'})(w_k-w_{k''})} +... ,
	\label{Eq.3a}
\end{align}
where $V_{k k'}= \langle k | V | k' \rangle$ and the cut-off $k_0$ ensures that all the momentum integrals involved in Eq. (\ref{Eq.3a}) are finite (see, e.g., \cite{wick_introduction_1955}, for more details).

Having solved $H_{\text{lab}}$, the next step is to examine $H_1$, the static Hamiltonian defined by taking into account mesons with momenta ranging over $]0, k_0[$ and $] \Lambda/2, \Lambda[$. The large scale-separation between these two slices ensures that low-energy mesons do not affect much the behavior of high-energy ones. We can thus divide $H_1$ into an ``unperturbed" high-energy Hamiltonian $H_0$ ranging over $]\Lambda/2, \Lambda[$ and a low-energy ``perturbation" $H_{\text{lab}}$ ranging over $]0, k_0[$. $H_{\text{lab}}$ has already been solved, at least approximately. $H_0$, on the other hand, is very similar to $H_{\text{lab}}$ except that it describes mesons with momenta of order $\Lambda$. The dependence on $\Lambda$ can be made explicit by rewriting $H_0$ as $\Lambda H_s$, with $ H_s$ the Hamiltonian obtained by rescaling the range of integration of $H_0$ down to $]1/2, 1[$, reparametrizing its creation and annihilation operators, and adjusting its parameters accordingly.

Note that the Hamiltonian $H_s$ includes interaction terms and does not have the same ground states $|p\rangle$ and $|n\rangle$ as the free part of the static model. Let us call $|P\rangle$ and $|N\rangle$ the physical ground states of $H_s$ with energy $E_0$. They are also the ground states of $H_0=\Lambda H_s$ with energy $\Lambda E_0$, i.e., the ground states of the bare nucleon source together with a cloud of virtual mesons of characteristic momentum $\Lambda$. Now, if we take into account $H_{\text{lab}}$, the contributions of low-energy ``laboratory" mesons are negligible compared to $\Lambda E_0$. We could, in effect, replace $|P\rangle$ and $|N\rangle$ by any state of the form $|P\rangle_l=|P, k_1, ..., k_l\rangle$ and $|N\rangle_m=|N, k_1, ..., k_m\rangle$ ($l, m > 0$) with low-energy mesons of momenta $k_i \ll \Lambda$ (but not too many to ensure that they indeed bring only small deviations at this level). That is: the physical ground state of $H_0$ is highly degenerate if we take into account low-energy laboratory mesons.

Wilson introduces his early prototype of effective Hamiltonian as a means to systematically evaluate how low-energy mesons affect the fixed source at some higher energy level. The key idea is the following. We could obtain the series of energy perturbations $E=E^{(0)}+ E^{(1)}+ E^{(2)}+...$ coming from low-energy mesons by applying degenerate perturbation theory to the high-energy Hamiltonian and using the series of perturbed states $|\psi\rangle=|\psi^{(0)}\rangle+ |\psi^{(1)}\rangle+|\psi^{(2)}\rangle+...$~. But this would become cumbersome if we have to repeat the analysis all over again each time we want to assess the impact of a different set of low-energy mesons. Instead, we can construct a new Hamiltonian which gives all the required energy perturbations when applied to the state-space spanned by the ground states $|\psi^{(0)}\rangle$ and adjust the corresponding basis whenever we want to add or remove low-energy mesons.

Consider first the level $\Lambda$. Wilson defines the effective Hamiltonian $H_{\text{eff}}$ at this level in the lowest order of degenerate perturbation theory by projecting $H_1$ onto the state-space spanned by the ground states $|P\rangle_l$ and $|N\rangle_m$ of its unperturbed part $H_0$. In this subspace, $H_0$ reduces to $ \Lambda E_0$. There is also a non-zero probability that the physical ground state of the fixed source shifts from $|N\rangle$ to $|P\rangle$ through the absorption of a low-energy meson of momentum $k$, i.e.,  $\langle P| a_k \tau^{+} | N, k \rangle=\alpha \neq 0$ for the transition amplitude. We can account for the effect of the cloud of mesons at the level $\Lambda$ by redefining the bare fixed-source operator $\tau^{+}$ in Eq. (\ref{Eq.1}) in terms of a renormalized operator $\tau_R^{+}= \tau^{+}/{\alpha}$, i.e., by using $|P\rangle=\tau_R^+ |N\rangle$ instead of $|p\rangle=\tau^+ |n\rangle$. As a result, the effective Hamiltonian takes the following form:
\begin{align}
H_{\text{eff}}= \Lambda E_0 +  & \frac{1}{(2\pi)^3} \int_{0}^{k_0}  d^3k \Big[w_k(a_k^{\dagger}a_k + b_k^{\dagger}b_k)\Big] \nonumber \\
&+ \frac{g_0\alpha}{(2\pi)^3}  \int_{0}^{k_0}  d^3k \Big[\frac{1}{\sqrt{2w_k}} \big[ (a_k+b_k^{\dagger})\tau_R^{+} +(a_k^{\dagger} +b_k)\tau_R^{-}\big] \Big] ,
\label{Eq.4}
\end{align}
with the same conventions as in Eq. (\ref{Eq.1}) and with a renormalized ``effective" coupling defined by $g_r= g_0\alpha$.\footnote{Note that if $H_{\text{eff}}$ is the projection of $H_1$ onto the state-space spanned by the states $|P\rangle_l$ and $|N\rangle_m$ ($l, m \geq 0$), we need to replace $\tau$ by $\tau_R$ in the expression of $H_{\text{eff}}$. Wilson corrects this minor point in his 1970 article (cf. \cite*[p. B449]{wilson_model_1965}; \cite*[p. A443]{wilson_model_1970}).}

Consider now the level $ \Lambda^n$. We can again divide the Hamiltonian $H_n$ ranging over all the slices up to $]\Lambda^n/2, \Lambda^n[$ into an unperturbed part describing mesons with momenta of order $\Lambda^n$ and a perturbative part describing mesons with momenta of order $\Lambda^{n-1}$ or less. The same type of effective Hamiltonian can be obtained from $H_n$ by projecting it onto the subspace spanned by the ground states of $\Lambda^n H_{s,n}$ together with mesons with momenta $\Lambda^{n-1}$ or less:
\begin{equation}
H_{\text{eff}, n}= \Lambda^n E_0[g_0] +  H_{n-1} [g_0 \alpha(g_0)] ,
\label{Eq.5}
\end{equation}
where the dependence on the bare coupling $g_0$ is made explicit. The form of $H_{n-1}$ is similar to the expression found in Eq. (\ref{Eq.4}), except that it takes into account slices up to the momentum scale $\Lambda^{n-1}$ and includes an operator $\tau_{R,n}$ defined relatively to the ground states $|P_n\rangle$ and $|N_n\rangle$ of $H_{0,n}$. 

Then, we can successively apply the transformation to the Hamiltonians $H_{m}$ ($m \leq n-1$), ignoring each time mesons with momenta of order $\Lambda^{m}$. This allows us to evaluate how the effects of low-energy mesons add up across scales and to obtain the lowest energy deviations of order $\Lambda^{m-1}$ above the lowest energy level of $H_{n}$:
\begin{align}
	&\Lambda^n E_0 [g_0] +  H_{n-1}, \nonumber \\
	&\Lambda^n E_0 [g_0] + \Lambda^{n-1} E_0 [g_0 \alpha(g_0)]+  H_{n-2}, \nonumber \\
	&  ..., \nonumber \\
	& \Lambda^n E_0 [g_0] + \Lambda^{n-1} E_0 [g_0 \alpha(g_0)] + ... + H_{\text{lab}} 
	\label{Eq.7}
\end{align}
The pattern is, in fact, similar to what is usually found in atomic physics: the recursive formula ultimately provides the ``maximally" hyperfine structure of the fixed source specified at the level $\Lambda^n$ given a series of effects organized according to their relative contributions in powers of $1/\Lambda$. 

The transformation also generates a sequence of couplings, going here from a ``bare" coupling $g_0$ at the level $\Lambda^n$ down to the ``renormalized" laboratory-level coupling $g$:
\begin{equation}
g_n=g_0 \; , \;  g_{n-1}= g_n \alpha(g_n) \; , \; ... \; , \;  g= g_1 \alpha(g_1)
\label{Eq.6}
\end{equation}
Although Wilson does not refer to it in this way, Eq. (\ref{Eq.6}) corresponds to his early formulation of an ``RG equation" in terms of a sequence of couplings at distinct energy scales. Each coupling $g_m$ ($0 < m <n$) in the sequence encodes effects arising from a cloud of virtual mesons at the level $\Lambda^{m+1}$ (and indirectly virtual effects up to the level $\Lambda^n$). Note as well that the RG equation takes a discretized form and that the ``renormalization scale" is specified by the slice number $m$.

The lowest energy levels of $H_n$ can also be found by starting with $ H_{\text{lab}}$ and some experimentally determined value of the coupling constant $g$ and inverting the recursion formula above. This amounts to studying the high-energy behavior of the static model by successively taking into account the effects of new high-energy slices on the physics originally described by $ H_{\text{lab}}$. And by investigating the behavior of $g_n$ and $H_n$ in the limit $n \to \infty$, one can obtain information about the eigenvalues of the ``butchered" static Hamiltonian and the strength of its coupling at high energies.

There are, of course, many aspects of this article that deserve further scrutiny. But we have reached a point where we have everything we need to assess the significance of Wilson's 1965 article with respect to his later works on the RG and EFTs. First, Wilson lays down two fundamental methodological principles that will become the trademark of his RG approach later on: (i) divide a complex problem involving many scales into a set of simpler sub-problems specified at different scales (multi-scale analysis); (ii) evaluate how the physics at some scale affects the physics at other scales (interscale-sensitivity analysis). Second, Wilson has a preliminary version of a ``coarse-graining" procedure: $H_{\text{eff}, n}$ is defined by ignoring ``high-energy" mesons with characteristic momentum $\Lambda^n$, keeping ``low-energy" mesons with momenta $k < \Lambda^{n-1}$, and taking into account the simplest effects of virtual high-energy mesons on the coupling between low-energy mesons and the fixed source. Third, Wilson's resulting prototype of EFT shares many elements with his version of the early 1970s. For instance, $H_{\text{eff}, n}$ represents a restricted set of degrees of freedom within a limited range of energy scales. The relative independence, or ``autonomy", between high-energy and low-energy degrees of freedom is also directly built into the mathematical formalism of the effective model (i.e., it is not something that we discover at the end of the renormalization procedure as in the traditional analysis of a renormalizable model). Finally, Wilson's early RG equation involves a discrete sequence of couplings at different \textit{physical} scales (and not, for instance, a sequence specified by arbitrary renormalization conditions). 

These similarities notwithstanding, Wilson is still far from his mature views. Note first that the high-energy structure of QFTs is still both a mathematically and physically relevant problem in Wilson's mind. For instance, he devotes the end of his 1965 article to classifying the possible asymptotic behaviors of couplings at high energies and still treats QED as a putatively UV-complete field theory.\footnote{``The curve A [i.e., the Landau pole singularity case] will never occur in practice. For if it did occur then $e_{\lambda}$ would not exist for large $\lambda$. But because of our definition of $e_{\lambda}$ [...], the photon propagator also would not exist for large photon momentum $k$. But quantum electrodynamics cannot exist without a photon propagator, and in particular the function $\phi$ [i.e., the beta function] will then not exist either. So, instead of the curve A being a possibility, we have the possibility that no consistent quantum electrodynamics exists." (\cite[p. B453]{wilson_model_1965})} Wilson is indeed far from suggesting the idea of treating realistic theories with a fixed cut-off at this stage. In his eyes, the main underlying goal is still to solve the full Hamiltonian of realistic theories and the renormalization procedure still requires taking the infinite cut-off limit at the end. Second, Wilson is not yet fully clear about the idea of integrating out high-energy degrees of freedom. $H_{\text{eff}, n}$ still describes the ``high-energy" system originally described by $H_n$ and not some new ``low-energy" system, strictly speaking. Third, non-renormalizable terms are absent in 1965. Wilson takes $H_{\text{eff}, n}$ to give the lowest-order term that one would obtain by solving $H_n$ in degenerate perturbation theory. The Hamiltonian $H_{\text{eff}, n}$ would be clearly more complicated if higher-order perturbative effects from low-energy mesons were included. But Wilson does not provide any detail at this stage and he is not yet explicit about the idea that $H_{\text{eff}, n}$ would include an infinite number of independent interaction terms. Finally, Wilson does not speak about irrelevant and relevant couplings and is not concerned with low-energy fixed-points in 1965.

Despite these missing elements, Wilson has pushed field theory forward in three major directions in 1965. First, he has turned the traditional way of parsing a theory on its head. The standard approach was to divide some Hamiltonian $H$ into a free and an interacting part and assume that the interacting part is sufficiently small to avoid intractable nonlinearities and derive approximate solutions. Wilson divides $H$ into a high-energy and a low-energy part and introduces a sufficiently large gap between the two to treat the low-energy part as a perturbation. Second, Wilson has provided a clear physical basis to the renormalization program and made it clear that the most important aspect of the renormalization procedure lies in the relationship between different physical scales and not in the high-energy limit of theories (see \cite[p. 4]{wilson_renormalization_1976}; \cite*[p. 590]{wilson_renormalization_1983}, for later remarks related to this point). Third, Wilson has developed a new type of field theory and thereby somewhat deviated from the pristine, albeit approximate, blueprint offered by QED. In particular, his notion of effective Hamiltonian in 1965 corresponds to a renormalized field theory with a sharp cut-off accounting both for the lowest energy level of a system specified at some higher energy scale and for small deviations induced by the physics at the cut-off scale and below.

Before assessing what were the main driving forces behind these achievements, I need to address one remaining puzzle: namely, that contrary to what Wilson (\cite*[p. B447]{wilson_model_1965}) says, the butchered static model does not display the same sorts of issues as realistic models and thus might not give us any interesting information about them. Take the issue of UV divergences for instance and suppose that for some scattering process, the original static model exhibits UV logarithmic divergences as in QED (i.e., $\int^{\Lambda}  dk/k \propto \ln (\Lambda)$ for $\Lambda \to \infty$). If we take into account mesons with arbitrarily high energies, the butchered static model exhibits linear divergences in the number $n$ of slices for the same process: 
\begin{equation}
 \sum_{m=1}^{n} \int^{\Lambda^{m}}_{\frac{1}{2} \Lambda^{m}} \frac{dk}{k} = n \ln 2 
 \label{Eq.7b}
\end{equation}
This suggests, first, that the high-energy structure of the butchered static model is different from that of realistic field theories and, second, that the methods developed here might not give any interesting information about realistic models. After all, it was already well-known at the time that if two models display different types of UV divergences, a method that works in one case might not work or be as fruitful in the other case. Renormalizable and non-renormalizable models provide a clear example with respect to standard perturbative methods (if one wishes to take the infinite cut-off limit). And even if the butchered static model were displaying the same kinds of divergences as realistic models, say, by identifying $n$ with $\ln(\Lambda/\mu)$, it would not be sufficient to conclude that their high-energy structure is similar and that they should be treated in the same way, as it is now clear with QED and Quantum Chromodynamics (QCD), the modern theory of the strong interaction.

In my view, the most plausible solution to this puzzle is to take Wilson to be mainly concerned with finding a new non-perturbative method and not discovering structural features of realistic models themselves. It would be surprising if he was not aware of the issues I just mentioned. For instance, although Wilson does not make it explicit in his discussion of the various asymptotic behaviors at high energies (\cite*[sec. VI-VII]{wilson_model_1965}), he is probably aware that the couplings of two models displaying the same kinds of UV divergences in perturbation theory may follow different trajectories at high energies. And if my reading is correct, Wilson is just not yet sure that the method he is developing in 1965 will work for realistic models. As we will see, he does spend the following years trying to make sure that he is not ``landing himself in the soup" with the sort of recursive transformation he develops in 1965. And he knows at the time that more work needs to be done to show that the method applies to realistic models. So, to put it differently, Wilson is not yet concerned with solving the general problem of how QFTs behave at high energies. He only means to make a more modest and cautious contribution: namely, to develop a method that \textit{might} help physicists make some progress in that direction.

\section{Taking stock}
\label{sec:Taking stock: What really made the difference?}

Now is a good time to take a pause and assess what really made the difference in the story so far. 

As we have seen, Wilson worked within different research traditions during his early career and came to be in contact with unusual ones for an aspirant field theorist. He first received the standard field-theoretic training of the 1950s and gradually learned that something new would have to be invented. His work in meson theory provided him with a simple, paradigmatic, and sufficiently realistic field-theoretic model to fiddle with and eventually brought him to the emerging RG tradition of the 1950s. Gell-Mann and Low gave him the idea of a sequence of couplings, though probably not the physical picture he developed in 1965. This picture rather came from the slicing method itself and became clearer to Wilson \textit{once} he compared it with Gell-Mann and Low's rather formalistic picture of scale-dependence (see \cite[p. B452]{wilson_model_1965}). Wilson was also in contact with the growing S-matrix tradition of the 1960s but quickly distanced himself from this more phenomenological approach to particle physics (see, e.g., \cite{chew_dubious_1963}). The model field-theoretic tradition of the 1950-60s provided him with more models to toy around with and cleared his mind for dealing with the static model in the way he deemed necessary. Van Hove is likely to be the main source of inspiration for his attempt to examine how a cut-off Hamiltonian model behaves when one successively adds new high-energy degrees of freedom to it. And there is more, from Wilson's interest in purely mathematical puzzles (e.g., \cite{wilson_proof_1962}) to his acquaintance with the field of quantum chemistry through discussions with his father, Bright Wilson (\cite{wilson_tribute_1993}).\footnote{Another potential source of inspiration behind Wilson's slicing method could be Freeman Dyson's idea of developing a renormalization technique based on the separation between low and high frequencies (\cite{dyson_renormalization_1951}). Wilson explicitly relates his own renormalization method to Dyson's idea in his later works (e.g., \cite[p. 1842]{wilson_renormalization_1971-1}; \cite*[footnote 14]{wilson_renormalization_1972}; \cite*[p. 249]{wilson_relativistically_1976}; \cite*[p. 590]{wilson_renormalization_1983}). At the same time, Wilson also acknowledges later on that the logic of his method is quite different from Dyson's (cf. \cite[1/4, 11'20''ff.]{wilson_interview_1991}; \cite*[part I]{wilson_interview_2002}).}

Despite all these influences, it does not seem that any of them brought Wilson to butcher the state-space of the static model and define a sequence of effective models in the specific way he did. For instance, Wilson's peculiar way of parsing a cut-off Hamiltonian into a high-energy and a low-energy part is remarkably distinct from Gell-Mann and Low's and his own early attempt to isolate the high-energy behavior of perturbative expressions. Wilson's infinite sequence of cut-off Hamiltonians with intermediary gaps of order $\Lambda$ is also quite different from Gell-Mann and Low's simple division between low-energy and high-energy scales (and from Dyson's division between low and high frequencies). Likewise, Wilson kept continuous slices in his 1965 article instead of following van Hove's suggestion to start directly with a fully discretized and finite model. But he \textit{did} rely on a discrete sequence of couplings (compared to Gell-Mann and Low). And overall, it seems that these sources of inspiration and the diversity of research traditions in which Wilson was working gave him at best new directions for his investigations and a strong incentive to try out new methods instead of using existing ones (as he originally did in his dissertation).

Wilson also worked on different local puzzles during the late 1950s and early 1960s. The first was to examine how the Low equation works with K-mesons, in line with Gell-Mann's interest in the phenomenological side of particle physics. This puzzle, however, quickly turned into the more mathematically-oriented one of understanding the asymptotic behavior of the solutions to the Low equation for any kind of meson. Then, in the following years, Wilson moved back and forth between specific mathematical puzzles related to the properties of phenomenological quantities (\cite{amati_theory_1963,wilson_regge_1963}) and the high-energy behavior of physically intuitive toy models (e.g., the Lee model, the Thirring model), ultimately moving back to field theory for good with his work on the operator product expansion (OPE) and model Hamiltonians in 1963-65. 

Again, Wilson did not follow a series of closely related puzzles that naturally brought him to the slicing method. At best, these puzzles were related insofar as they helped him to get a grip on the more general problem of understanding the structure of QFTs. As Wilson recalls: ``I knew that I wanted a problem that I could work on for a long time" (\cite*[part 2, 4'43''ff.]{wilson_interview_2010}). He did not, however, engage with this problem by showing undue reverence for the pristine formulation of field theory inherited from QED compared, say, to physicists attempting to apply gauge symmetries to the weak and strong interactions (e.g., \cite{schwinger_theory_1957,salam_weak_1959}). Wilson's return to field theory in 1963 was rather pragmatic. On the one hand, he thought that understanding its structure was the mathematically most interesting problem in physics he could work on at the time. This explains why he had no qualms about twisting the structure of field theory if it meant gaining new insights. On the other hand, Wilson thought that the best way of understanding the structure of field theory was to develop new non-perturbative methods and he took this project to be mathematically interesting in itself. This explains why he worked with simple models that might not have the relevant UV behavior and why he made diverse and somewhat unrelated attempts at probing their high-energy structure. And overall, it seems that the local puzzles Wilson worked on during this period played a rather secondary role compared to these more stable and long-lasting pragmatic aspirations. 

But was it enough? Consider the role played by Wilson's concern with the UV together with his more pragmatic aspirations in the context of his 1965 article. Wilson knew that it was essential to have a good handle on the UV behavior of $H_{\text{static}}$ in order to solve it. Traditionally, the problem involved two parts: (i) analyze the different types of divergences displayed by the model of interest; (ii) assess whether the model is perturbatively renormalizable. Wilson made an analogous move: (i) analyze how the high-energy degrees of freedom of the model affect its low-energy behavior; (ii) assess its limiting UV behavior. Since $H_{\text{static}}$ displayed UV divergences, Wilson had to introduce a cut-off in order to address these questions. And since $H_{\text{static}}$ was non-relativistic and unrealistic at high energies, there was no reason for him not to use the simplest one, namely, a sharp momentum cut-off $\Lambda$. To address (i), it thus makes sense that Wilson would divide a cut-off Hamiltonian into a high-energy and a low-energy part, which is equivalent to adding high-energy degrees of freedom to some original low-energy model.

Now, the crucial question for Wilson was: how would this analysis help to solve $H_{\text{static}}$? Wilson knew that high-energy mesons had more important effects than low-energy ones. It would thus make sense to treat the low-energy Hamiltonian as a ``perturbation" to the high-energy one. But this new kind of perturbative analysis would work only if there was a sufficiently large scale-separation between their energy levels. This explains why Wilson inserted a large energy gap of characteristic size $\Lambda$ between the ``low-energy" and ``high-energy" systems described by $H_{\text{lab}}$ and $H_0$, respectively. We can also understand why Wilson introduced a first prototype of effective theory at all. It was meant as a tool to assess how the low-energy physics described by $H_{\text{lab}}$ affects the lowest energy levels of $H_0$, which, again, is equivalent to evaluating the impact of adding new high-energy degrees of freedom to $H_{\text{lab}}$.

We still face two issues: (i) Why did Wilson take into account an infinite number of slices? (ii) Why did he define his prototype of effective theory over a continuous momentum range? We might try to address these issues by pointing out that Wilson's specific way of analyzing the impact of high-energy degrees of freedom forced him to rely on a sequence of cut-off Hamiltonians if he also wanted to evaluate the limiting UV behavior of the model. After all, $H_0$ was defined over the range $]\Lambda/2,\Lambda[$ and it was not obvious whether taking the limit $\Lambda \to \infty$ would give any relevant information. On the other hand, $H_{\text{eff}}$ and $H_{\text{lab}}$ were, as a matter of design, restricted to low-energies. So Wilson was forced to re-conceptualize the UV behavior of the model in terms of a discrete number $n$ of high-energy slices and consider an infinite sequence by taking the limit $n \to \infty$. This reconstruction somewhat helps to address (i). But it does not explain why Wilson felt the need to keep continuous slices. He could have directly worked with lattice field-theoretic methods by introducing a large-distance and a short-distance cut-off.

The crucial missing element is Wilson's general outlook about how to solve physical problems and thus make decisive progress in physics at the time. Wilson had been fascinated since an early age by the ability of machines to solve problems thanks to his grandfather who was a ``gear expert" at MIT (\cite*[p. 5]{wilson_supercomputers_1983}; see also \cite*{wilson_letter_1986}). Wilson became increasingly interested in computers during his studies at Caltech through his interactions with Jon Mathews, who introduced him to elementary programming problems with the computers of the Jet Propulsion Lab (\cite[p. 5]{alexander_american_1982}; \cite[part I]{wilson_interview_2002}). A few years later, Wilson tried to use the MIT computers to gain some insight into his dissertation work. But they had too little computing power for Wilson to get anything interesting out of them within a reasonable amount of time and he got increasingly frustrated by this experience (\cite*[part II]{wilson_interview_2002}). Yet this failure proved to be quite productive. Wilson was led to develop a new methodological scheme: the physical problems he worked on would have to be framed in such a way that they could be solved in principle by a sufficiently powerful computer (\cite*[p. 5]{wilson_supercomputers_1983}; \cite*[part II]{wilson_interview_2002}). And this new methodological scheme played a crucial role in the way he thought about butchering the static model after his early work on the OPE (\cite*{wilson_products_1964}):
\begin{quote} 
	So then I spent a long time just trying to figure out how do I get more understanding about high-energy behavior [...]. The perturbative approach doesn't work. So that sets me off thinking about a computational approach to go beyond perturbation theory. And I particularly get interested in the question if I had a computer big enough, how would I do this computation. That became sort of a guiding point to drive me to how to think about the problem. How can I convert this from a problem of infinite number of degrees of freedom, which you can't deal with anyhow, to a problem which is finite even if it was so large that you would have to have an astronomical size computer. I just wanted to convert it from an infinite number of degrees of freedom to a finite number. (\cite[part I]{wilson_interview_2002})
\end{quote}

As we have seen, Wilson did not directly convert the problem into a finite one in his 1965 article. Each slice still contained an infinite number of degrees of freedom. But Wilson did something physically more subtle and methodologically more simple. He implemented the foundational methodological principle at the basis of any sensible algorithmic strategy, the so-called ``divide and conquer" principle: (i) he ``divided" a complex and seemingly unsolvable problem involving a continuous and infinite range of momenta into an infinite series of simpler sub-problems of the same type involving a continuous yet limited range; (ii) he ``combined" the solutions of these sub-problems to solve the original problem. Each Hamiltonian $H_{0,n}$ ranging over $]\Lambda^n/2, \Lambda^n[$, including $H_{\text{lab}}$ at the laboratory level, was indeed conceived from the get-go as a tractable copy of the original Hamiltonian $H_{\text{static}}$. Both $H_{0,n}$ and $H_{\text{static}}$ involved an infinite number of degrees of freedom for instance. But $H_{0,n}$ was at least approximately solvable compared to $H_{\text{static}}$. Of course, implementing the division was not enough. Wilson also needed to justify that these sub-problems could be analyzed separately before relating them to one another---hence the assumption that the slices are sufficiently separated across energy scales. And by making this assumption, Wilson found that the infinite number of sub-problems would reduce to a two-fold recursive problem: (i) solve the lowest-energy Hamiltonian $H_{\text{lab}}$, i.e., the inductive base; (ii) derive the recursive relationship between $H_n$ and $H_{n+1}$, which involves, in particular, solving $H_{s,n+1}$ and expressing $g_{n+1}$ in terms of $g_n$.

There is another benefit coming out of this: namely, that this perspective helps to understand why the main irony in Wilson's part of the story is harmless. As we have seen, a key part in Wilson's original project was to understand the high-energy structure of QFTs. And yet he ultimately came up with a new type of field theory, which, as a matter of principle, could not provide such information. If I had only singled out Wilson's concern about the UV and not his pragmatic aspirations and methodological commitments, his early formulation of EFTs would have looked somewhat accidental at this stage. Likewise, if I had only singled out the traditions in which he was working at the time, it would have been hard to explain why Wilson defined an infinite sequence of low-energy effective models (instead of following Gell-Mann and Low's method for instance). Wilson's path to his first prototype of EFT and the specific form it takes only make sense if we acknowledge his pragmatic aspirations and peculiar algorithmic outlook on physical problems.

\section{The road ahead and the meaning of field theory}
\label{sec:The road ahead and the meaning of field theory}

Wilson continued to work on model Hamiltonians and on his early formulation of the OPE in the following years. He eventually published two major articles in 1969-70 and reached his mature conception of the RG and EFTs in 1971 (\cite{wilson_non-lagrangian_1969,wilson_model_1970,wilson_renormalization_1971,wilson_renormalization_1971-2}). Along the way, Wilson's concern for the high-energy behavior of QFTs became less central than it originally was in the late 1950s and early 1960s. He also became clearer about the necessity of including all possible terms in an EFT, the idea of integrating out high-energy degrees of freedom, and the significance of low-energy fixed-points, relevant and irrelevant operators, and large scale-separation. Wilson's algorithmic outlook kept playing an important role in these transformations (see, esp., \cite{wilson_model_1970}). But he also made new encounters, including a decisive one with statistical physics in 1965-66. Since this episode has already received a fair amount of attention in the literature (e.g., \cite{cao_conceptual_1993,schweber_hacking_2015}), I will be much briefer in what follows.

So we are in 1965, at the time Wilson has published his first article on the RG and EFTs. The article will receive literally no attention in the following years, and Wilson is, in fact, the first and only physicist who cites it in the early 1970s (cf. \cite{wilson_model_1970,wilson_renormalization_1971-2,wilson_renormalization_1972}). He also finds little success at conferences. For instance, Gerald Guralnik recalls that Wilson ``got beaten up rather badly" when he presented his ideas about discretized model Hamiltonians in July 1965 (\cite[p. 21]{guralnik_history_2009}). Yet Wilson's 1965 work appears to have a deep impact on his subsequent ideas and lead him, in particular, to rethink in depth the meaning of field theory (see, e.g., \cite{wilson_letter_1965}; \cite*[sec. 7]{wilson_renormalization_1971-1}; \cite[sec. 1.1]{wilson_renormalization_1974}). As he recalls later on in his Nobel Prize lecture:
\begin{quote}
	Following this development [the 1965 article], I thought very hard about the question ``what is a field theory," using the $\phi^4$ interaction of a scalar field [...] as an example. [...] I realized that I had to think about the degrees of freedom that make up a field theory. (\cite[pp. 590-1]{wilson_renormalization_1983})
\end{quote}
At the time, the tradition in particle physics, especially among those who work with Lagrangian models, is to think about ``degrees of freedom" in terms of the different kinds of particles, whether elementary or composite, that constitute a given physical system, and distinguish between them by attributing appropriate quantum numbers and symmetry transformation properties to their corresponding field variables (e.g., \cite{schwinger_theory_1957,schwinger_field_1964,glashow_partial-symmetries_1961,salam_electromagnetic_1964}).

Wilson seems to take a diametrically opposite view. Like many other model field theorists, he has worked for many years with simple model Hamiltonians describing essentially only one kind of field (e.g., a meson field for the static model). Wilson has also relied heavily on their formulation in terms of creation and annihilation operators, which makes their structure across energy scales more explicit. And when he publishes his article on model Hamiltonians in 1965, the important question is not whether a field is composite or elementary, for instance, or whether it has appropriate mass or spin properties to account for existing particle phenomena. Rather, it seems that the central issue at the heart of field theory for Wilson is to understand how distinct degrees of freedom contribute across different physical scales---at first in momentum space in 1965, and eventually in position space by the mid-1970s, when he starts to devote most of his time to lattice field theory. It was common at the time to regard the Fourier decomposition of a field in momentum space as a convenient formal trick and take creation and annihilation operators as bookkeeping devices for tracking changes in the occupation states of the field (e.g., \cite[pp. 89-92]{schweber_mesons_1955}; \cite[sec. 6.e]{schweber_introduction_1961}). In his 1965 article, Wilson rather takes the decomposition in momentum space to be constitutive of the meaning of a field as a multi-scale system and thinks of it in terms of a sequence of sub-systems across physical scales.

Wilson also appears to become increasingly aware around this time that he needs to introduce a cut-off in order to escape the pitfalls of perturbation theory and understand the structure of realistic QFTs (as he also emphasizes later on in \cite*[p. 591]{wilson_renormalization_1983}). Not every kind of cut-off will do, however. As we have seen, Wilson needs a cut-off that is simple enough to separate unambiguously different contributions according to their relative importance across scales---hence Wilson's use of a sharp cut-off in momentum space. The ``phase-space cells" method he develops in 1965 may appear by today's lights to be a rather confusing attempt at separating wave packets (see \cite[pp. 164-72]{wilson_construction_1965}; \cite*[sec. VII]{wilson_model_1965}; \cite*{wilson_renormalization_1971-2}). But the strategy is the same. In both cases, Wilson picks out one degree of freedom at a time, or one ``pack" of degrees of freedom, according to its relevance at a particular scale. And by using the simplest type of cut-off, a sharp cut-off, Wilson is very close to working with a lattice field theory, as he does later on.

It is important to emphasize that Wilson's outlook is still mainly methodological at this stage. In his eyes, the slicing and phase-space cells methods offer a conceptually clear and efficient way of probing the structure of QFT (see, e.g., \cite[p. B447]{wilson_model_1965}; \cite*[p. 591]{wilson_renormalization_1983}). But he does not believe that realistic QFTs only make sense with a fixed cut-off. Wilson takes QED to be a putatively UV-complete theory, as we saw above, and he is also open to the possibility that some field-theoretic models might display a high-energy fixed-point (see, e.g., \cite*[sec. VI]{wilson_model_1965}). In the same vein, his 1965 article indicates that he attributes more and more importance to the question of how different degrees of freedom are related to one another across scales compared to his original concerns about the asymptotic behavior of field theories at high energies. But again, the shift is largely methodological at this stage. Wilson's theoretical tools seem to underwrite a new vision of field theory which might prove to be remarkably efficient for solving traditional puzzles. He still needs, however, to test whether these tools---and thus this new vision---are fruitful with the help of new toy models, as he has done before for the behavior of QFTs at high energies. And it is precisely at this moment that statistical physics enters the stage.

\section{Wilson's encounter with statistical physics}
\label{sec:Wilson's encounter with statistical physics}

\begin{quote} 
	[...] the land of statistical physics is broad, with many dales, hills, valleys and peaks to explore that are of relevance to the real world and to our ways of thinking about it. (\cite[p. 127]{fisher_renormalization_1999})
\end{quote}

\noindent Wilson had not been very much in touch with or interested in statistical physics when he first came to Cornell in 1963 (\cite[part I]{wilson_interview_2002}). In 1965-66, however, a series of fortuitous encounters with physicists working on critical phenomena led him to redesign in depth his research plans, perfect his new field-theoretic tools, and reach what is now considered to be his mature understanding of the RG and EFTs. I will briefly outline Wilson's three most significant encounters in what follows (see \cite{schweber_hacking_2015}, for more details).

The ``first catalytic event", in Wilson's own words, dates back to 1965-66, before he went to the Aspen Center for Physics in Colorado in the summer 1966 (\cite[part II]{wilson_interview_2002}). Benjamin Widom, who had just been appointed full professor in the Chemistry Department at Cornell in 1963, gave a talk on his new scaling hypothesis and Wilson turned out to be there, though probably by chance.\footnote{There is some ambiguity about the exact date of this talk. Williams (\cite*[p. 8]{williams_historical_2016}) briefly mentions Wilson's own hesitations about the exact date, around 1964-65 according to his interview with the Physics of Scale project (\cite[part I]{wilson_interview_2002}) and ``sometimes before going to Aspen" according to his Nobel Prize lecture (\cite[p. 591]{wilson_renormalization_1983}). Widom places the date either in 1966 or 1967 (\cite[part III]{widom_interview_2003}). It seems to me that Wilson probably heard Widom's talk either toward the end of 1965 or early in the year 1966 since Widom was on leave at the University of Reading during the first semester of 1965 and Wilson started to work on problems in statistical physics when he went to Aspen.} Widom (\cite*{widom_equation_1965}) had found a simplified equation of state for a fluid at criticality. This equation would involve a homogeneous function $\phi(\lambda x, \lambda y)= \lambda^{\gamma} \phi(x,y)$ of its variables $x$ and $y$, with $\gamma$ some degree of homogeneity and $\lambda$ some scaling factor, and thus display a simple (multiplicative) scaling behavior with no natural dimensionful scale near some critical point $(x^*,y^*)$. Wilson (\cite*[1/4, 33'48''ff.]{wilson_interview_1991}) recalls finding the result puzzling. He had encountered similar simple asymptotic behaviors for systems displaying multiple scales in his previous works. Yet Widom had not provided any justification for his scaling equation apart from its empirical adequacy, and this moved Wilson to explore further the connection between particle physics and critical phenomena.

The second catalytic event for Wilson was Leo Kadanoff's 1966 preprint introducing the idea of a ``blocking transformation". Following Widom's clarification of the scaling behavior of a fluid near criticality, Kadanoff (\cite*{kadanoff_scaling_1966}) took up the project of devising a new strategy for justifying Widom's (\cite*{widom_equation_1965,widom_surface_1965}) results. He started with the Ising model and decided to divide the lattice into blocks smaller than the average correlation length between the spins, viz. over distance scales where the spins would tend to line up and effectively act as small ``blocks". As we will see in more detail below, Kadanoff replaced the spin variables contained within each block by a single one and adjusted the parameters of the Hamiltonian to account for the effects of the missing spins. According to Kadanoff, the resulting effective Hamiltonian would have the same form as the original one close enough to the critical point. And by performing this blocking transformation, he was able to obtain a set of functional equations for some key quantities of the model and show that they scale according to Widom's homogeneity hypothesis (\cite[sec. 3]{kadanoff_scaling_1966}). 

There is, in fact, some uncertainty as to whether Wilson had worked out a blocking transformation before reading Kadanoff's preprint. Wilson (\cite*[p. 591]{wilson_renormalization_1983}) suggests in his Nobel Prize lecture that he was working on similar ideas in Aspen and discovered that he ``had been scooped" after discussing his work with some condensed matter physicists there, who directed him toward Kadanoff's preprint. But it is also possible that Wilson was doing something quite different in Aspen and encountered Kadanoff's preprint only afterward, during one of the seminars organized by Benjamin Widom and Michael Fisher at Cornell (\cite[pp. 84-5]{mermin_early_2015}). Be that as it may, Wilson's later works and remarks still suggest that Kadanoff's work helped him to understand more clearly the origin of Widom's scaling laws and the connection between condensed matter and particle physics (e.g., \cite{wilson_renormalization_1971}; \cite*[part II]{wilson_interview_2002}; \cite[p. 82]{wilson_renormalization_1974}). In both cases, physical problems would involve multiple scales which do not all participate in the same way in the phenomena of interest. Likewise, Kadanoff's work probably helped him to clarify the idea of ``integrating out" physically irrelevant degrees of freedom and realize that the physically important notion of fixed-point is a low-energy one, not a high-energy one (\cite[p. 82]{wilson_renormalization_1974}; \cite[part II]{wilson_interview_2002}). 

The third catalytic event was Wilson's encounter with Fisher who had just moved from King's College to Cornell in the summer 1966. The two would soon deliver groundbreaking work, both relevant to the fields of condensed matter and particle physics (\cite{wilson_critical_1972}). At the beginning, however, Wilson found in Fisher a privileged source of wisdom to catch up with the recent developments in statistical physics (\cite[1/4, 45'22''ff.]{wilson_interview_1991}). Fisher was already a well-established figure in the field, having published important articles and reviews about critical phenomena (e.g., \cite{fisher_susceptibility_1959,fisher_correlation_1964,fisher_theory_1967,essam_pade_1963}). He also had a far-reaching understanding of what was at stake. Thanks to him, Wilson discovered that statistical physics was full of ideas similar to the ones he had already been toying around with in the past few years, including ideas about effective Hamiltonians, scaling laws, and non-integral scaling exponents. And from there, as Wilson puts it in his Nobel Prize lecture, he started to amalgamate his ``thinking about field theories on a lattice and critical phenomena" (\cite*[p. 591]{wilson_renormalization_1983}).

The first article he published after 1965 with a graduate student at Cornell, Kenneth Piech, displayed such amalgamation (\cite{piech_model_1968}). The field-theoretic model they introduced to account for atomic phenomena was directly formulated on a lattice. Inversely, Wilson learned from Fisher and Widom that the Ising model he had been working on since 1966 could serve again as ``a `theorists' laboratory' for the investigation of the same kind of phenomena that lead to renormalizability challenges in quantum field theory" (\cite[p. 4]{wilson_origins_2005}). Wilson had thus expanded his portfolio of model field theories. But now it seems that their status had somewhat evolved in his mind. They had stopped being mere toy models to be studied for the sake of obtaining information about the high-energy behavior of realistic theories in particle physics. Wilson had also found that they were, to a certain extent, physically realized in some condensed matter systems and could thus be treated as autonomous field-theoretic models at least in some areas.

\section{``I wanted to have that assurance that I wasn't just landing myself in the soup"}
\label{sec:``I wanted to have that assurance that I wasn't just landing myself in the soup"}

Another fundamental element in Wilson's mind changed around 1966-67. It became clear to him that the exact version of the RG transformation he had worked out in 1965 would generate an infinite number of independent terms parametrized by distinct couplings, namely, all the possible terms consistent with the symmetries of the original model he had started with. At the same time, he also realized that the new terms would be systematically organized according to their importance across scales and that most of them would become ``irrelevant" at the scale of interest. 

The origin of this shift in Wilson's mind is not entirely clear, however. One plausible hypothesis is that he was already aware of the need to introduce all the possible terms thanks to his work on the OPE in 1963-64. Wilson's (\cite*{wilson_products_1964}) manuscript contains a similar methodological principle. The product of local operators evaluated at points arbitrarily close to one another is expanded in a complete basis of arbitrarily complex local operators. As far as I can tell, this is the first time Wilson thinks of field operators as operator-valued vectors living in an infinite-dimensional vector space. And he is also aware that the set of operators in this basis is not only fully determined by the symmetries and variables of the model of interest but also includes in principle both perturbatively renormalizable and non-renormalizable operators.

Now, Wilson does introduce an additional constraint on the dimensionality of these operators, which, in effect, is akin to the constraint of perturbative renormalizability often endorsed at the time (\cite*[pp. 27-8, 32]{wilson_products_1964}). This constraint reflects his interest in the singular behavior of QFTs at short distances, which is primarily contained in the first few lowest-dimensional terms of an OPE in standard cases. And yet, despite this somewhat arbitrary restriction, Wilson still endorses a methodological principle akin to what is known today as Gell-Mann's ``totalitarian principle", which states that everything that is not forbidden is compulsory (cf. \cite[p. 859]{gell-mann_interpretation_1956}). Wilson indeed emphasizes that an OPE needs to include all possible terms that are consistent with a given set of symmetries and constraints (e.g., \cite*[pp. 28-9, 32]{wilson_products_1964}). And this may well have opened his eyes about the need to include all the possible terms consistent with the principles of a given model when one successively eliminates one scale after the other in a physical problem.

The connection is nonetheless too thin to be significant. First, Wilson was concerned in 1964 with the behavior of products of operators at distinct points, not about the most general form of a local field-theoretic operator, say, the Hamiltonian density $H(x)$ of a given model. There is, in other words, no reason for him to think that a scale-dependent transformation on a local operator (as he does it in 1965) automatically generates all the local operators satisfying the same principles as the original one. Second, Wilson's manuscript suggests that the kind of systematicity he is after in 1964 is more akin to the systematicity sought when including higher-order terms in the perturbative calculation of a given correlation function. That is, he is not making a claim about the most general form of models themselves but rather trying to give a non-perturbative generalization of the renormalization procedure.\footnote{See Wilson (\cite*[pp. 4-5, 11, 25]{wilson_products_1964}) for his suggestion that the OPE might provide the basis for a new non-perturbative renormalization program.}

A better hypothesis, in my sense, is that Wilson's attitude shifted after he read Kadanoff's work in 1966. Although Wilson was already ``moving in that direction", as he emphasizes later on (\cite*[part II]{wilson_interview_2002}), Kadanoff probably gave him a clear set-up to examine more precisely how the dynamics of a model changes when one eliminates one scale of a physical problem at a time. As Wilson further recalls:
\begin{quote} 
	[Kadanoff] just postulates, of course, that decimation would come back to the same Hamiltonian, and one of the things I do work through at some point is doing decimation on two-dimensional Hamiltonian, but which, of course, produces infinite sets of couplings, not just two as Kadanoff assumed. (\cite[part II]{wilson_interview_2002})
\end{quote}
In other words, Wilson convinced himself that a typical RG transformation would generate all the possible terms consistent with the principles of some original model by fiddling with the two-dimensional Ising model. I have not come across archival materials where Wilson does this calculation explicitly. In particular, his manuscripts on the Ising model in the 1960s do not include such calculation.\footnote{See Wilson's ``Unpublished Notes on the Ising Model" dated from the 1960s in Kenneth G. Wilson Papers, \#14-22-4086. Division of Rare and Manuscript Collections, Cornell University Library. Box 1, Folder 6.} But if his recollections in his interview with the Physics of Scale project are correct, he probably did the calculation in 1966-67 (i.e., after he arrived in Aspen and before he did a similar calculation with the static model in 1967 or so, cf. below).

\begin{figure}[h!]
	\centering
	\begin{center}
		\includegraphics[scale=0.8]{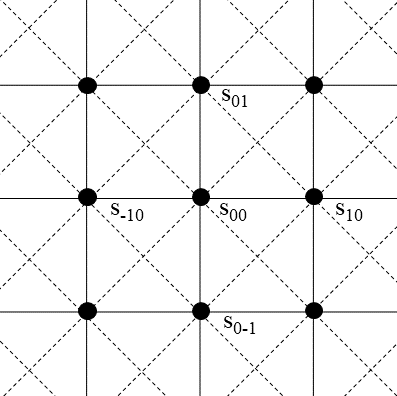}
		\caption{\small{Decimation of one diagonal over two in the two-dimensional Ising model.}}
		\label{fig1}
	\end{center}
\end{figure}

The best way to understand Wilson's shift is to briefly show how a simple blocking transformation generates new terms in the Ising model.\footnote{I will follow Yeomans's simple presentation here (\cite*[sec. 9.2]{yeomans_statistical_1992}). But more details can be found in Goldenfeld (\cite*[sec. 9.6]{goldenfeld_lectures_1992}) and Kadanoff (\cite*[sec. 14.2]{kadanoff_statistical_2000}) for instance.} Consider first a two-dimensional lattice of infinite size and spacing $a$ with a two-valued spin variable $s_i=\pm 1$ located at each site $i$. The dynamics of the system is governed by a simple Hamiltonian: 
\begin{equation}
H=J \sum_{\langle ij\rangle} s_is_j , 
\label{Eq.8}
\end{equation}
where each spin $s_i$ interacts only with its nearest neighbors $s_j$ on the lattice. For simplicity, I have set the usual term $h \sum_i s_i $ accounting for the interaction between the spins and an external field $h$ to zero. Then, the next step is to implement a simple decimation procedure eliminating one diagonal of spins over two in the lattice (see Fig. \ref{fig1}). The procedure is equivalent to eliminating only the spin $s_{00}$ at the center of each block made of five spins $s_{00}$, $s_{01}$, $s_{0-1}$, $s_{10}$ and $s_{-10}$ obtained after rotating the lattice by $45$ degrees (using a simplified notation $s_{ij}$ to keep track of the interactions between the five spins below). The result is a lattice with spacing $a\sqrt{2}$ having exactly the same shape as the original one. 

At the formal level, the decimation procedure goes as follows. Note first that the original partition function,
\begin{equation}
	Z=\sum_{\langle s\rangle}\exp{\Big[-H(\langle s \rangle)\Big]}, 
	\label{Eq.9}
\end{equation}
with $\langle s\rangle$ a possible lattice spin configuration, can be rearranged as a sum of products of terms involving only the five spins of each rotated block:
\begin{equation}
\exp{\Big[-J s_{00}(s_{01}+ s_{0-1} +s_{10} + s_{-10})\Big]}
\label{Eq.10}
\end{equation}
If we eliminate the spin $s_{00}$ at the center of each block by explicitly computing the partition function for each $s_{00}=\pm 1$, we obtain a sum of products of terms like:
\begin{equation}
2\cosh{\Big[-J (s_{01}+ s_{0-1} +s_{10} + s_{-10})\Big]}
\label{Eq.11}
\end{equation}
By considering the sixteen possible configurations ${s_{ij}=\pm1}$, this expression can be rewritten as: 
\begin{align}
\exp \Big[-A(J) - B(J)( &s_{-10} s_{01} + s_{01} s_{10}+ s_{10} s_{0-1}+ s_{0-1} s_{-10}   \nonumber\\
	& + s_{-10} s_{10}+ s_{0-1} s_{01}) - C(J)(s_{-10} s_{01} s_{10} s_{0-1})\Big] ,
\label{Eq.12}
\end{align}
provided the new parameters $A(J)$, $B(J)$, and $C(J)$ satisfy:
\begin{align}
&A(J)=\ln2 + \big[\ln \cosh (4J) + 4 \ln \cosh (2J)\big]/8 \nonumber\\
&B(J)=\big[\ln \cosh (4J)\big]/8 \nonumber\\
&C(J) = \big[\ln \cosh (4J) - 4 \ln \cosh (2J)\big]/8 
\label{Eq.13}
\end{align}
We can thus eliminate the spin $s_{00}$ in each diagonal block and obtain a thermodynamically equivalent system by taking into account the interaction between every pair of the remaining four spins as well as their collective interaction. Ignoring the constant term, the new effective Hamiltonian is given by:
\begin{equation}
H_{\text{eff}}= 2 B(J)  \sum_{\langle ij\rangle} s_is_j +B(J)  \sum_{[ij]} s_is_j +C(J)  \sum_{(ijkl)} s_is_js_ks_l ,
\label{Eq.14}
\end{equation}
with $\langle ij\rangle$ indexing nearest neighbors in each new block, $[ij]$ the two spins lying across each other, and $(ijkl)$ the four spins of the block. The transformation has thus generated two new types of interaction terms. Each iteration introduces new terms encoding the effects of interactions involving the spins that have been decimated. And if the procedure is iterated infinitely many times, the resulting effective Hamiltonian includes infinitely many distinct types of interaction terms.

Of course, finding such a pattern in one paradigmatic example was far from proving that it would obtain in every case. Yet Wilson belongs to this category of physicists for whom having two distinct paradigmatic examples satisfying one general pattern is enough to convince oneself that it applies in many cases (cf. \cite*[4/4, 20'12''ff.]{wilson_interview_1991}). And he needed to check whether the static model displayed the same pattern. The difficulty, in fact, was not so much in finding out whether the effective Hamiltonian at the level $\Lambda^n$ obtained by taking into account higher-order perturbations from low-energy mesons included infinitely many terms. Wilson was probably already aware of that (cf. \cite*[p. 592]{wilson_renormalization_1983}; \cite*[p. 7]{wilson_origins_2005}). He had not worked out the complete structure of $H_{\text{eff},n}$ in 1965. But he still had defined it as the Hamiltonian that would give all the required higher-order corrections in the state-space spanned by the ground states of the system, which would otherwise have been obtained by applying degenerate perturbation theory to $H_n$ and using perturbed states. For this to work, $H_{\text{eff},n}$ had to include higher-order operators of the schematic form:
\begin{equation}
	V \frac{1}{E^{(0)} - H_0} V ... V \frac{1}{E^{(0)} - H_0} V ,
	\label{Eq.15}
\end{equation}
with $H_0$ the unperturbed Hamiltonian, $E^{(0)}$ its eigenvalues, and $V$ the perturbation (i.e., respectively, $\Lambda^n H_s$, $\Lambda^n E_0$, and $H_{n-1}$ in the notation of section \ref{sec:The slicing method in 1965}). The real difficulty, then, was to assess whether the resulting expansion and the transformation between $H_{\text{eff},n}$ and $H_{\text{eff},n-1}$ made any mathematical sense.

Wilson spent the next few years making sure that he was indeed not ``landing [himself] in the soup by just imagining these Hamiltonians with [an] infinite number of couplings" (\cite*[part I]{wilson_interview_2002}; see also \cite*[p. 4]{wilson_renormalization_1976}; \cite*{wilson_origins_2004}; \cite*[p. 13]{wilson_origins_2005}). According to his recollections, he did most of the work in 1967 or so (\cite*[cf.][4/4, 19'47''ff.]{wilson_interview_1991}). The final product, however, arrived only three years later in a thirty-five-page-long article titled ``Model of Coupling Constant Renormalization" (\cite{wilson_model_1970}). Wilson was able to show in this article that the mathematical space in which these Hamiltonians live is well-defined and closed under the transformations relating them. He thus obtained the confirmation from his study of the static model that both Gell-Mann and Low's and Kadanoff's original pictures were too simple to be correct (as he recounts in \cite*[p. 6]{wilson_renormalization_1975-1}).
 
The key element that convinced Wilson that he was not landing himself in the soup is the existence, in today's terms, of a systematic power-counting scheme for evaluating the contributions coming from different scales. Wilson was indeed concerned about whether an RG transformation would ultimately involve only a manageable subset of couplings (\cite*[part III]{wilson_interview_2002}). He realized in the case of the static model that he could easily separate the different contributions coming from low-energy mesons according to their importance across scales and show that increasingly complex contributions are increasingly negligible at the scale of interest (\cite[sec. III]{wilson_model_1970}). Wilson actually provides in this article a much more rigorous analysis than this simple picture, explaining why the contributions of new interaction terms do not increase without bounds when one includes increasingly many higher-order perturbative terms.\footnote{See also Wilson's ``Unpublished Notes on Topological Analysis---Convergence of Operator Interaction in a Renormalization Group Transformation in the Fixed Source Problem" in Kenneth G. Wilson Papers, \#14-22-4086. Division of Rare and Manuscript Collections, Cornell University Library. Box 1, Folder 10.} It will not be necessary to go into the details here. The important point is that higher-order perturbative terms obtained at the scale $\Lambda^{n-1}$ contribute to the value of the energy levels at the scale $\Lambda^n$ by increasing powers of $1/\Lambda$ (cf. Eq. \ref{Eq.15} above). Most of the new terms, in other words, are irrelevant at the scale of interest.  Wilson was led, in turn, to study in more detail the contributions of what he later designated as relevant and irrelevant operators and realize that there was no need to introduce a large scale-separation between different slices.\footnote{Kadanoff had already been using the term `irrelevant variable' as Wilson acknowledges in (\cite*[p. 3179, footnote 11]{wilson_renormalization_1971}). See also Wilson and Kogut (\cite*[p. 112]{wilson_renormalization_1974}).} The large energy gaps introduced in 1965 had only allowed him to safely ignore higher-order perturbations from one scale to the other. But he could have taken more perturbations into account and examined how two scales arbitrarily close to one another affect each other. 

\section{The state of affairs in 1971}
\label{sec:The state of affairs in 1971}

We are now in 1971, at a point where all the essential elements of Wilson's mature conception of the RG and EFTs are on the table. Starting with a bare model with a sharp cut-off $\Lambda_0$, a full and non-perturbative RG transformation generates a series of effective models restricted to lower scales $\Lambda < \Lambda_0$ and including infinitely many independent interaction terms, organized according to the importance of their contributions at the scale $\Lambda$. In typical cases, the transformation eventually hits a low-energy fixed-point around which only a finite number of interaction terms remain relevant. And given this achievement, we might wonder whether Wilson still takes the high-energy limit of QFTs to be a physically relevant problem at this point. I will briefly conclude the story by providing some details about the two syntheses of 1970 and 1971 and Wilson's views toward the UV in 1971.

The first synthesis comes with Wilson's 1970 article ``Model of Coupling Constant Renormalization" already discussed above. He has simplified his early version of the static model and now works with a lattice model with only one degree of freedom within each slice. The idea of integrating out high-energy degrees of freedom is also completely clear by now. Non-renormalizable terms have also become perfectly acceptable in Wilson's mind (cf. \cite*[pp. 1440, 1446]{wilson_model_1970}). But most importantly, Wilson lays down for the first time the modern apparatus of his version of the RG and specifies, in particular, the concept of theory-space and RG transformation in formal terms (see \cite*[esp. sec. V]{wilson_model_1970}). Within this setting, the properties of a low-energy effective model generated by an RG transformation are more determined by the transformation itself and the existence of a low-energy fixed-point than by the properties of the original bare model (\cite*[pp. 1456, 1460]{wilson_model_1970}). That is, the RG appears to be \textit{constitutive} of what an effective theory is in Wilson's 1970 article. Wilson, however, is not attributing any explicit physical significance to cut-offs here. They are mainly introduced for mathematical reasons, i.e., to efficiently derive an RG transformation and thus understand the structure of the model across scales. The question of the UV limit is also clearly not a concern for Wilson in this article. He wants to ``understand renormalization better" (\cite[p. 1439]{wilson_model_1970}). And renormalization itself, not the ``problem of renormalization" associated with the definition of a renormalized QFT in the UV limit, is understood here in terms of the variational structure of a theory across two different scales.

Wilson does not stop there. The second synthesis comes in 1971 with his article ``Renormalization group and Critical Phenomena II: Phase-Space Cell Analysis of Critical Behavior", which unfolds the picture laid down in 1965 and 1970. Wilson has already provided an intuitive topological picture of the RG in a first short article by drawing an analogy with the behavior of classical flows in phase-space (\cite*{wilson_renormalization_1971}). The RG space is, in Wilson's words, carved by ``hills", ``gullies" and ``ridges", and simple topological explanations are provided to account for the behavior of RG flows. Wilson's second article provides the mathematical and methodological counterparts of this picture (\cite*{wilson_renormalization_1971-2}). This is the first time Wilson integrates out high-energy degrees of freedom by eliminating variables lying on external ``momentum shells" and obtains arbitrarily complex interaction terms as a result---a method very much akin to what is now done in the path integral formalism.\footnote{Wilson explicitly makes the connection in his lectures on the RG at Cornell in the early 1970s (cf. \cite*[p. 31]{wilson_unpublished_1972}).} He is thus able to provide a new mathematical basis for Kadanoff's somewhat arbitrary blocking transformations and a clear physical picture explaining why integrating out high-energy degrees of freedom generates infinitely many interaction terms. The new terms encode the complex set of correlations involving both the remaining degrees of freedom and those that have been eliminated.

Wilson's increasing skepticism toward the physical relevance of the high-energy limit of QFTs becomes clearer in an article published in the same year (\cite*{wilson_renormalization_1971-1}). This article is the occasion for Wilson to review Gell-Mann and Low's early work on the RG and examine the various possible high-energy behaviors of QFT models. Wilson explicitly acknowledges the ``academic nature" of such studies (\cite*[p. 1818]{wilson_renormalization_1971-1}). He is also probably largely convinced like many other physicists at the time that realistic field theories do not display a UV fixed-point (although he does mention the possibility of a limit cycle at high energies for the strong interaction in this article). Yet the key element in my sense rather lies in the set of reasons he suggests for keeping a cut-off fixed. Some of these reasons are of course not new. Wilson is also far from taking any of them to be sufficient. He is not even taking a definitive stance toward the idea of keeping a cut-off fixed. But it is striking to find him explicitly laying down for the first time some of the reasons that are usually mentioned today to support the idea that a given theory is best treated as an effective theory. 

First, and most importantly, Wilson raises the issue of treating a given interaction in isolation (\cite*[pp. 1819, 1830, 1832]{wilson_renormalization_1971-1}). A model might display a UV fixed-point and appear to be well-behaved at high energies. But if the model fails to account for some interaction that affects the physical processes it describes at high energies, there is little chance that it will remain reliable in these regimes. According to Wilson, it is physically sensible to introduce a fixed cut-off $\Lambda$ in such cases:
\begin{quote} 
	While the renormalized coupling constant $e^2$ of electrodynamics is small, one sees from Fig. 3 that $e_{\lambda}^2 \to x_1$ for $\lambda \to \infty$ [with $e_{\lambda}$ the renormalized coupling specified at some arbitrary scale $\lambda$ and $x_1$ some hypothetical UV fixed-point]. The constant $x_1$ is fixed independent of $e^2$ and so cannot be arbitrarily small. This suggests that all particles will couple strongly to photons at sufficiently high momenta; but this would mean that electrodynamics and strong interactions would mix strongly, suggesting that pure electrodynamics is valid only below a cutoff momentum $\Lambda$. (\cite[p. 1832]{wilson_renormalization_1971-1})
\end{quote}

Second, Wilson mentions the issue of the perturbative breakdown of a given model at high energies (\cite*[p. 1838]{wilson_renormalization_1971-1}). If the perturbative coupling of a model becomes increasingly large at high energies, there is no reason to believe that perturbation theory is reliable in those regimes. In the absence of non-perturbative methods, it is therefore methodologically sensible to restrict the model to regimes where it is perturbatively reliable by introducing a fixed cut-off. 

Finally, Wilson emphasizes the idea that there is no fundamental distinction between (perturbatively) renormalizable and non-renormalizable interaction terms contrary to the standard lore at the time (\cite*[p. 1840]{wilson_renormalization_1971-1}). Non-renormalizable terms may well be required in some circumstances. As we will see in Weinberg's part of the story, chiral Lagrangians provide a good example. And if a model includes non-renormalizable terms, it is essential to keep the cut-off fixed to ensure that the renormalized model is predictive beyond tree-level. 

So, although the UV limit of QFTs is still a mathematically interesting problem in Wilson's eyes, it appears that he has come to take less seriously its potential physical relevance compared to his early views in the 1960s. Of course, more needs to be said about the reasons behind this shift, including Wilson's missed opportunity with asymptotic freedom. The important point for now is that he has not only laid down his mature conception of EFTs in the early 1970s but also explicitly put forward a set of preliminary justifications for treating a theory as an EFT.

\section{Conclusion}
\label{sec:conclusion}

Kenneth Wilson's path to effective field theories (EFTs) was marked by his long-standing aspiration to understand the high-energy structure of putatively UV-complete field theories. Wilson was first brought to study the high-energy behavior of a simple meson-nucleon model, the static model, and encountered through this work the early renormalization group (RG) and model field-theoretic traditions of the 1950-60s. After a quick detour through the S-matrix program and a first attempt with the operator product expansion, he eventually came to divide the static model into an infinite sequence of effective models describing phenomena at distinct energy scales. Before anything else, Wilson's first prototype of EFT was the result of his attempt to break down an intractable problem into smaller and more manageable pieces by drawing energy scales apart. 

His subsequent encounter with statistical physics eventually led him to clarify how these models are exactly related to each other across scales, thereby giving a physically more intuitive and mathematically more rigorous basis to the set of renormalization methods that had proved so successful with the electromagnetic interaction. By the early 1970s, Wilson's mature conception of the RG and EFTs was in place. In particular, Wilson's works made it clear that the structure of an EFT is largely determined by an RG transformation together with a low-energy fixed-point. Both the mathematical framework and the physical picture he had developed since 1965 quickly drew many adepts, especially in the condensed matter physics community (see, e.g., Domb's short historical account of Wilson's impact in \cite*[sec. 1.7]{domb_critical_1996}). As Shankar puts it: ``It was love at first bite. [...] I seriously considered adding a middle initial G to my name to openly flaunt my newfound faith." (\cite*[p. 204]{shankar_rg_2015})

I will briefly end Wilson's part of the story by mentioning two outstanding ironies. We have already seen that Wilson's pragmatic aspirations and algorithmic methodological outlook explain why he was led to formulate a prototype of EFT which, on the face of it, could not give him any information about the high-energy behavior of QFTs. In the same vein, it is also somewhat ironic that Wilson gained further physical insights into the hierarchical structure of matter across scales from the theory of critical phenomena. Critical phenomena are, after all, exactly the sorts of phenomena for which every scale equally matters, which may well explain why Kadanoff thought he had to obtain the same description at large distance scales. By letting non-renormalizable terms in, Wilson ended up with a radically different physical picture where the physics the most relevant at some scale becomes completely irrelevant at some other scale. This apparent irony, however, disappears once we realize that Wilson was inspired only by Kadanoff's strategy of going from one scale to the other and did not make any assumption about how the structure of physical phenomena would vary across these scales. 

There is yet one last irony for which there is not much to be done. Wilson spent more than ten years looking for non-standard perturbative methods and non-perturbative methods to understand the high-energy behavior of the strong interaction when it turned out to be unnecessary. The discovery of asymptotic freedom in 1972-73 led physicists to realize that the high-energy regime of the strong interaction is precisely the regime for which non-perturbative methods are not needed. It is hard to imagine how Wilson could have anticipated this discovery, viz. that the strong coupling regime of the strong interaction lies at low energies and not at high energies. Be that as it may, his enduring efforts to treat QFTs in non-perturbative terms still proved to be a remarkably fruitful way of studying QCD at low energies.

\medskip

{\small 
	\noindent \textbf{Acknowledgments:} I would like to thank Alexander Blum and Porter Williams for invaluable archival materials and very fruitful discussions. I am also grateful to Michael Peskin, James Fraser, Vincent Ardourel, audiences at the MPIWG, 2nd ESHS Early Career Scholars Conference and Foundations 2021, as well as the referees, for very useful exchanges and/or feedback. This work was supported by the Max Planck Institute for the History of Science.
}

%\nolinenumbers

%\begingroup
%\setstretch{1}
%\setlength\bibitemsep{0pt}
\printbibliography
\end{document}